\documentclass{aa}  

\usepackage{graphicx}

\usepackage{txfonts}
\begin{document} 

   \title{A comprehensive comparison between APOGEE and LAMOST}

   \subtitle{Radial Velocities and Atmospheric Stellar Parameters}

   \author{B. Anguiano,
          \inst{1,2}
          S. R. Majewski,
          \inst{1}
          C. Allende-Prieto, 
          \inst{3,4}
          S. Meszaros,
           \inst{5,6} 
           H. J\"onsson,
           \inst{7} 
           D. A. Garc\'ia-Hern\'andez,
            \inst{3,4} 
           R. L. Beaton, 
          \inst{8,9} 
          \fnmsep\thanks{Hubble Fellow}
          \fnmsep\thanks{Carnegie-Princeton Fellow}
          G. S. Stringfellow,
           \inst{10}
          K. Cunha,
          \inst{11,12}
          V. V. Smith
          \inst{13}
          }

   \institute{Department of Astronomy, University of Virginia, Charlottesville, VA 22904-4325, USA \\
    \email{ba7t@virginia.edu}
            \and
           Department of Physics \& Astronomy, Macquarie University, Balaclava Rd, NSW 2109, Australia 
            \and
             Instituto de Astrof'sica de Canarias (IAC), E-38205 La Laguna, Tenerife, Spain
             \and 
             Universidad de La Laguna (ULL), Departamento de Astrof'sica, E-38206 La Laguna, Tenerife, Spain
             \and
             ELTE E\"otv\"os Lorand University, Goth\'ard Astrophysical Observatory, Szombathely, Hungary
             \and
            Premium Postdoctoral Fellow of the Hungarian Academy of Sciences
            \and
            Lund Observatory, Department of Astronomy and Theoretical Physics, Lund University, Box 43, SE-22100 Lund, Sweden
            \and
            Department of Astrophysical Sciences, Princeton University, 4 Ivy Lane, Princeton, NJ~08544
            \and
            The Observatories of the Carnegie Institution for Science, 813 Santa Barbara Street, Pasadena, California 91101, USA
            \and
            Center for Astrophysics and Space Astronomy, University of Colorado, 389 UCB, Boulder, CO 80309-0389, USA
            \and
            University of Arizona, Tucson, AZ 85719, USA
            \and
	   Observat\'orio Nacional, S\~ao Crist\'ov\~o, Rio de Janeiro, Brazil 
	   \and
	   National Optical Astronomy Observatories, Tucson, AZ 85719 USA\\}
             

   \date{Received May 7, 2018; accepted Jul 18, 2018}

 
  \abstract
   {}
   {We undertake a critical and comprehensive comparison of the radial velocities and the main stellar atmospheric parameters for stars in common between the latest data releases from the APOGEE and the LAMOST surveys.}
   {APOGEE is a high-resolution, high-signal-to-noise ratio spectroscopic survey, part of the SDSS. The latest data release, SDSS DR14, comprises APOGEE spectra for 263,444 stars, together with main stellar parameters and individual abundances for up to 20 chemical species. LAMOST is a low-resolution optical spectroscopic survey, where LAMOST DR3 contains 3,177,995 stars.}
   {There is a total of 42,420 dwarfs/giants stars in common between the APOGEE DR14 - LAMOST DR3 stellar catalogs. A comparison between APOGEE and LAMOST RVs shows a offset of 4.54 $\pm$ 0.03 km s$^{-1}$, with a dispersion of 5.8 km s$^{-1}$, in the sense that APOGEE RVs are larger. We observe a small offset in the effective temperatures of about 13 K, with a scatter of 155 K. A small offset in [Fe/H] of about 0.06 dex together with a scatter of 0.13 dex is also observed. We notice that the largest offset between the surveys occurs in the surface gravities. Using only surface gravities in calibrated red giants from APOGEE DR14, where there are 24,074 stars in common, a deviation of 0.14 dex is found with substantial scatter (0.25 dex). There are 17,482 red giant stars in common between APOGEE DR14 and those in LAMOST tied to APOGEE DR12 via \emph{the Cannon}. There is generally good agreement between the two data-sets. However, we find dependencies of the differences of the stellar parameters on effective temperature. For metal-rich stars, a different trend for the [Fe/H] discrepancies is found. Surprisingly, we see no correlation between the internal APOGEE DR14 - DR12 differences in T$_{\rm eff}$ and those in DR14 - LAMOST tied to DR12, where a correlation should be expected since LAMOST has been calibrated to APOGEE DR12. We also find no correlation between the [Fe/H] discrepancies, suggesting that LAMOST/Cannon is not well coupled to the APOGEE DR12 stellar parameters scale. An [Fe/H] dependence between the stellar parameters in APOGEE DR12 and those in DR14 is reported. We find a weak correlation in the differences between APOGEE DR14 - DR12 and LAMOST on DR12 surface gravity for stars hotter than 4800 K and in the log g range between 2.0 and 2.8 dex. We do not observe an [Fe/H] dependency in the gravity discrepancies.}
   {}


\titlerunning{A comprehensive comparison between APOGEE and LAMOST}
\authorrunning{Anguiano et al.}

\maketitle

\section{Introduction}

The structure of a stellar atmosphere is principally determined by three atmospheric parameters; the effective temperature, the surface gravity and the chemical enrichment level. These parameters are the foundation of the physical interpretation of stellar spectra and, together with radial velocities, the most important measurements sought after for the different scientific goals of current massive stellar spectroscopic surveys. Hence automated stellar parameter pipelines (SPPs) and their validation are extremely important for the scientific exploitation of these survey data \cite{2008AJ....136.2050L,2011AJ....141...89S,2015MNRAS.451.1229A}.

In this study we focus on the latest data releases from two large ongoing observational programs, the APOGEE \cite{2017AJ....154...94M} and LAMOST \cite{2012RAA....12..723Z} surveys. Our main goal is to work out a comprehensive comparison between the radial velocities and the main stellar atmosphere parameters for the stars in common between the two surveys. This comparison can illuminate how similar the stellar velocity and atmospheric parameters are for each survey, to evaluate random errors, and to find systematic effects within the data-sets. This comparison is especially helpful because LAMOST observes in the optical at medium-low resolution, while APOGEE survey works in the $H$-band at higher resolving power, and they use different analysis methodologies to obtain the information from the observed spectra.  

Data-driven methods for measuring stellar parameters, like \emph{The Cannon} \cite{2015ApJ...808...16N}, have been developed in the recent years to bring surveys with disjoint wavelength coverage and different resolving power onto the same scale using a training sample observed by both spectroscopic programs. Recently, \cite{2017ApJ...836....5H} used \emph{The Cannon} to transfer information from APOGEE Data Release 12 (DR12) to determine precise stellar parameters from the spectra of 450,000 LAMOST giants. In this study we also use the intersection between the APOGEE and LAMOST data-sets to compare the more recent Data Release 14 (DR14) of APOGEE to the LAMOST giants calibrated to the APOGEE DR12 scale. Moreover, asteroseismology in the \emph{Kepler} field \cite{2010Sci...327..977B} provides accurate surface gravities for red giants. Using the overlap between APOGEE and LAMOST, and also the photometric survey SAGA \cite{2014ApJ...787..110C} in the \emph{Kepler} field, we are able to quantify the discrepancies between surface gravity derived from spectroscopic analysis and those from asteroseismology \cite{2014ApJS..215...19P}. 

This paper is organized as follows. In Section 2 we describe the APOGEE and LAMOST surveys together with the stellar catalog of common stars. Sky coverage, magnitude range, signal-to-noise ratio, radial velocities and stellar parameters together with their uncertainties and systematic effects are discussed in detail. The APOGEE and LAMOST targets in the \emph{Kepler} field are described in Section 3. We emphasize that in Section 2 and 3 we compare APOGEE with LAMOST's own pipelines. In Section 4 we study the discrepancies between APOGEE and LAMOST calibrated to the APOGEE DR12 stellar parameters scale via \emph{The Cannon}. We present our conclusions and how this massive overlap between the surveys presents an opportunity to built a robust training data-set for data-driven methods for measuring stellar parameters in Section 5.

\section{The APOGEE-LAMOST stellar catalog}

The Apache Point Observatory Galaxy Evolution Experiment (APOGEE), and its successor APOGEE-2, is a high-resolution (R $\sim$ 22,500), high-signal-to-noise (SNR $>$ 100 per half resolution element) spectroscopic survey using the 2.5-meter Sloan telescope in the Northern Hemisphere and the du Pont telescope at Las Campanas Observatory for the Southern Hemisphere \cite{2017AJ....154...94M}. The survey operates in the near-infrared $H$-band, and can take 300 spectra simultaneously \cite{2010SPIE.7735E..1CW}. The latest APOGEE data release, DR14 \cite{2018ApJS..235...42A}, comprises spectra for 263,444 stars, together with main stellar parameters and individual abundances for up to 15 chemical species.  

The Large sky Area Multi-Object Spectroscopic Telescope (LAMOST) is a national scientific research facility operated by the Chinese Academy of Sciences. LAMOST is a low-resolution (R $\sim$ 1800), optical (3650 - 9000 \AA) spectroscopic survey in the Northern Hemisphere. Using a modified Schmidt telescope, LAMOST can observe up to 4000 objects simultaneously over a 20 sq. deg field-of-view. The LAMOST Experiment for Galactic Understanding and Exploration (LEGUE) is an on-going Galactic survey with a present sample of more than 5 million stellar spectra \cite{2012RAA....12..735D}. LAMOST DR3\footnote{http://dr3.lamost.org/} published 3,177,995 stars in this catalog, including 45,826 A type stars, 988,947 F type stars, 1,600,512 G type stars and 542,710 K type stars. These objects are selected with the criteria of having a SNR in the $g$ band larger than 6 obtained during dark nights, and SNR in $g$ band larger than 15 obtained during bright nights.

Using a comparison of positions in equatorial coordinates between the surveys, we selected stars where ($\Delta\alpha^{2}$ + $\Delta\delta^{2}$)$^{1/2}$ $<$ 3 arcsec, and find a total of 42,420 stars in common between APOGEE DR14 and LAMOST DR3. For this study we used the APOGEE flag called $ASPCAPFLAG$, to remove stars where any of the $TEFF, LOGG, CHI2, COLORTE, ROTATION, SNR$ flags are set (see \cite{2015AJ....150..148H} and \cite{2016AJ....151..144G} for full details on the different flags used in APOGEE data). This leaves 41,547 objects in the APOGEE - LAMOST stellar catalog.

\subsection{Sky coverage and magnitude range}

\begin{figure}
  \centering  
  \includegraphics[width=1.\columnwidth]{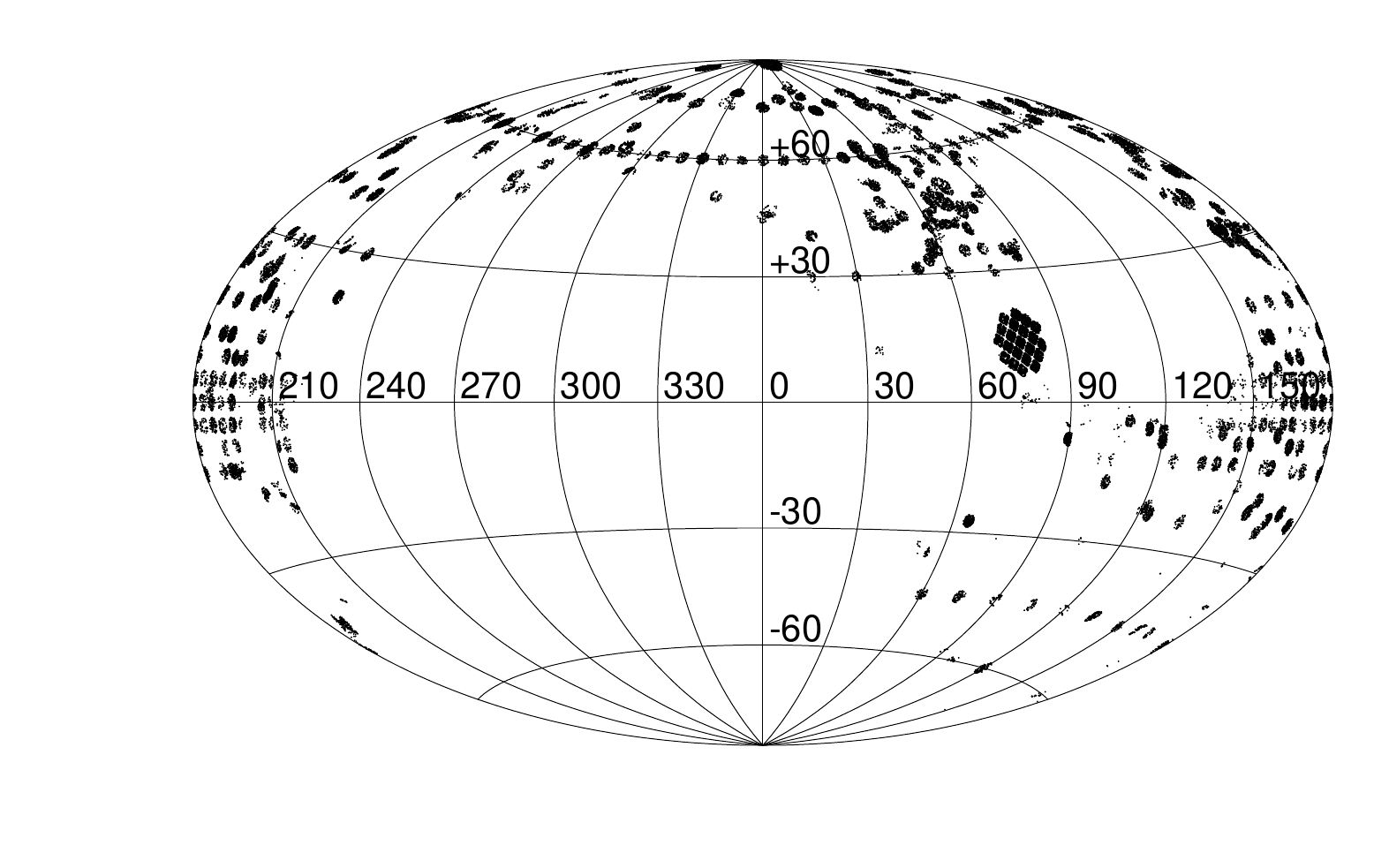}
   \caption{A Hammer-Aitoff  projection in Galactic coordinates (l,b) of the APOGEE-LAMOST stellar catalog distribution. Most of the stars are in the North Galactic Pole and in the Galactic anti-center. Note also the \emph{Kepler} field around $l$ $\sim$ 70$^{\circ}$, $b$ $\sim$ +15$^{\circ}$}
  \label{fig:sky}
\end{figure}

In Figure~\ref{fig:sky} we show the Galactic coordinates of the APOGEE-LAMOST stellar catalog in a Hammer-Aitoff projection. Because of the Northern Hemisphere location of LAMOST and the SDSS-telescope APOGEE used for DR12 and DR14, most of the common stars lie in the Galactic anti-center and in the North Galactic Cap.  

There are also common targets in the \emph{Kepler} field, where the spectroscopic parameters provided by the APOGEE project are complemented with asteroseismic surface gravities, masses, radii, and mean densities determined by members of the Kepler Asteroseismology Science Consortium (KASC) \cite{2014ApJS..215...19P}. There are similar efforts within the LAMOST collaboration in the \emph{Kepler} fields \cite{2016ApJS..225...28R}. 

\begin{figure}
  \centering  
  \includegraphics[width=1.\columnwidth]{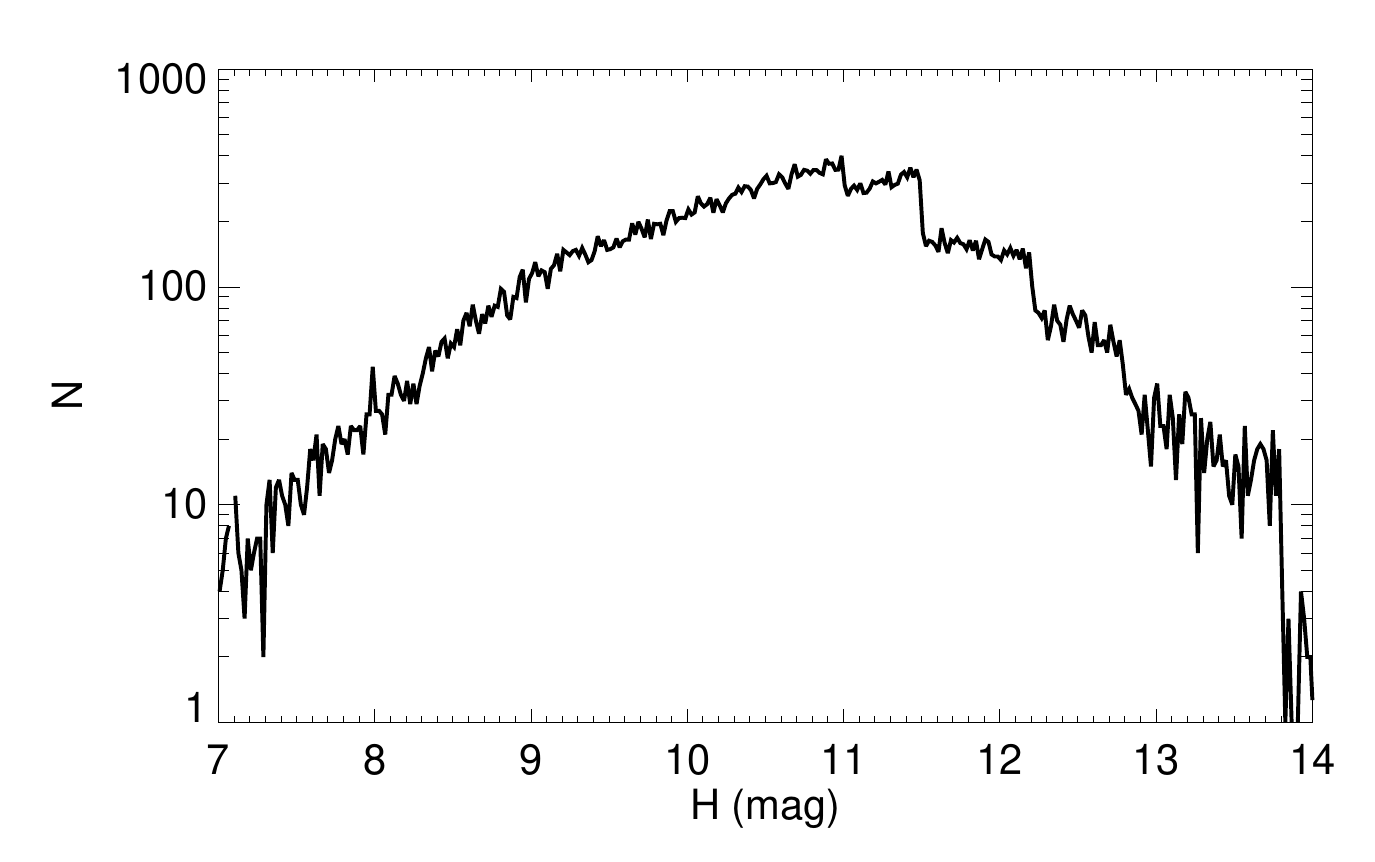}
   \caption{Luminosity distribution function in the $H$-band for the APOGEE - LAMOST stellar catalog. The catalog covers a large magnitude range from 7 to almost 14 mag. Most of the stars are in the magnitude range from 9 to 12 in the $H$-band.}
  \label{fig:mag_APOLA}
\end{figure}

Figure~\ref{fig:mag_APOLA} shows the $H$-band luminosity distribution for the APOGEE - LAMOST stellar catalog compiled in this study. Most stars in common between APOGEE and LAMOST are in the magnitude range from 9 to 12 in the $H$-band. 

\subsection{Signal-to-noise ratio (SNR)}

\begin{figure*}
  \centering  
  \includegraphics[width=1.05\columnwidth]{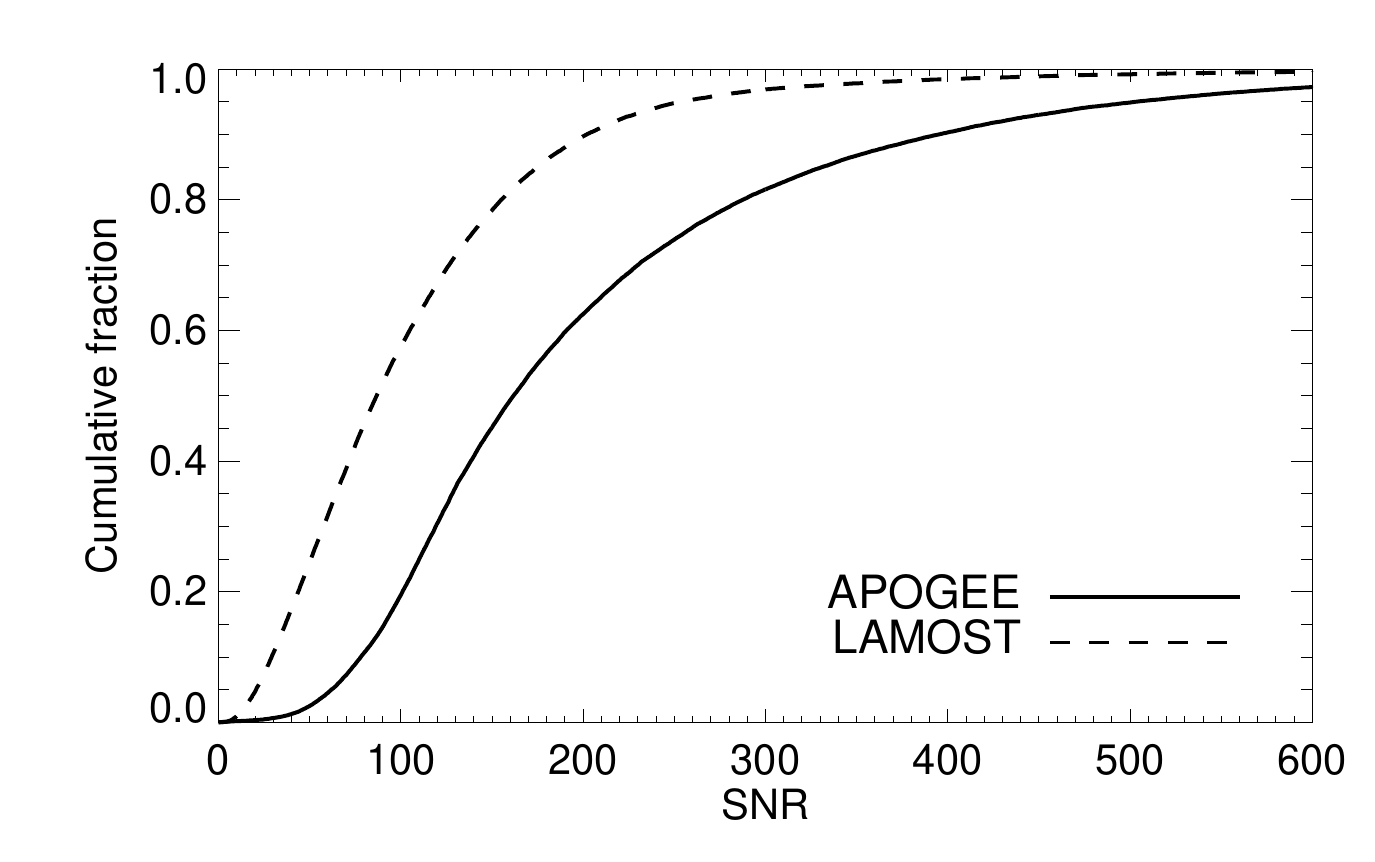}
   \includegraphics[width=0.9\columnwidth]{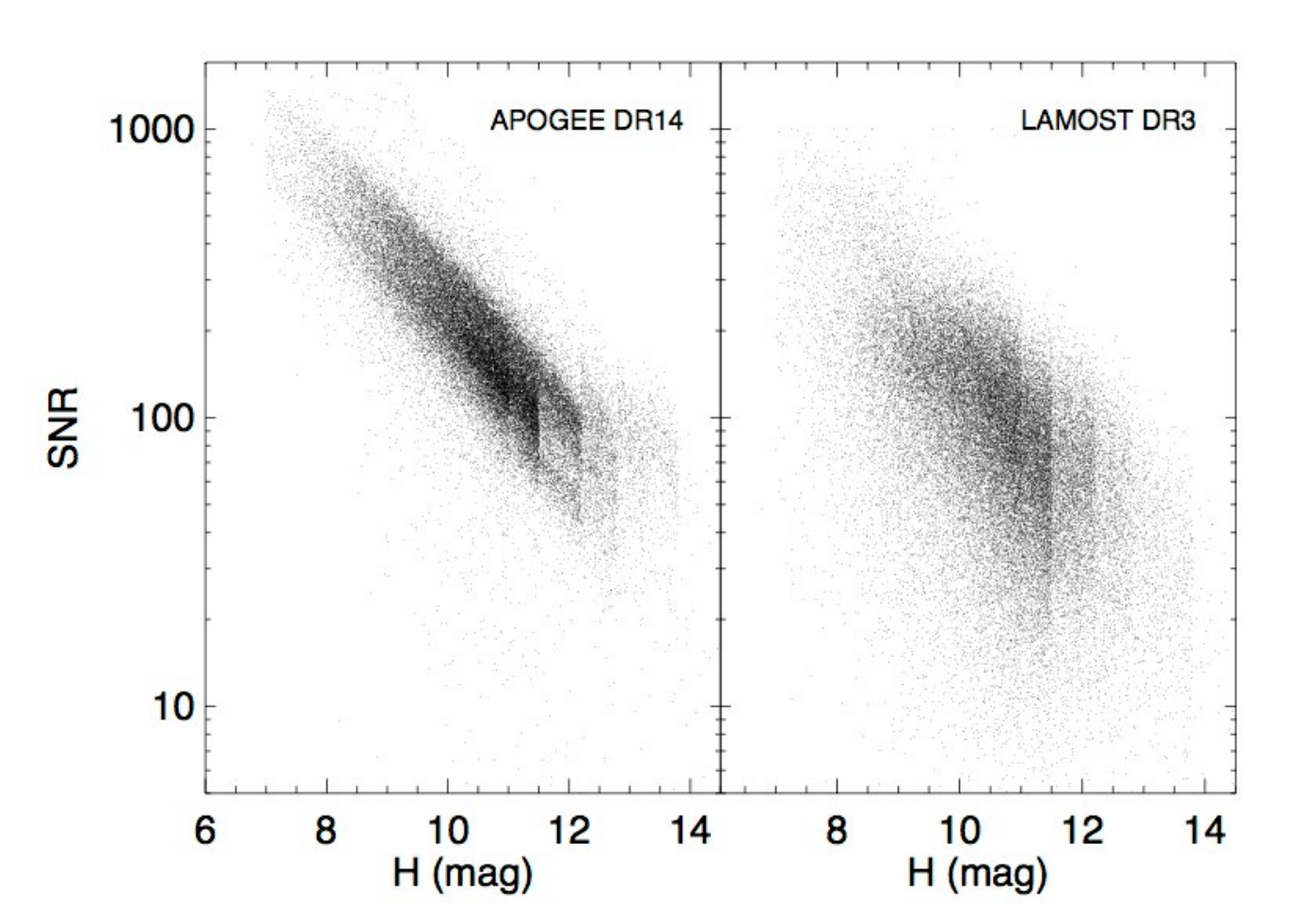}
   \caption{Left panel: cumulative histograms of SNR-per-half resolution element (approximately a pixel) for APOGEE (solid line) and SNR-per-pixel for LAMOST (dashed line), as provided from the original catalogs. Right panels: SNR from APOGEE and LAMOST spectra for the common objects as a function of the $H$-band magnitude from the 2MASS survey.}
  \label{fig:SNR_APOLA}
\end{figure*}

APOGEE aims to obtain high-resolution, high SNR ($>$ 100 per half resolution element of pixels), $H$-band spectra for stars in the bulge, disk, and halo of the Milky Way \cite{2013AJ....146...81Z,2017arXiv170800155Z}. Most APOGEE targets are observed in several visits. After multiple visits, a combined spectrum is produced \cite{2015AJ....150..173N}. LAMOST DR3 provides the SNR-per-pixel for the targets as calculated in five different bands, $u$, $g$, $r$, $i$ and $z$, respectively. Using the center wavelength and bandwidth, they obtain the wavelength range for each SDSS band, and then the SNR in each band is the median value at each pixel in this band \cite{2015RAA....15.1095L}. The cumulative histograms of SNR in APOGEE and in LAMOST for the $z$-band (left panel in Fig.~\ref{fig:SNR_APOLA}) shows that the APOGEE survey has 80$\%$ of the in common stars with SNR $>$ 100, and 40$\%$ with SNR $>$ 200. The LAMOST survey has 40$\%$ of the in common stars with SNR $>$ 100, and 10$\%$ with a SNR $>$ 200\footnote{APOGEE SNR is not real for SNR $>$ 200 - 300 since it is obtained from the photon shot noise. At very high SNR it becomes limited by the detector flatfielding, artifacts and telluric/sky lines.}. The right panel in Figure~\ref{fig:SNR_APOLA} shows the SNR from APOGEE and LAMOST spectra for the common stellar catalog as a function of the $H$-band magnitude from the 2MASS survey \cite{2006AJ....131.1163S}. This figure shows a clear relation between magnitude and SNR in APOGEE. This relation is weaker in LAMOST DR3 data and the $H$-band, where a given stellar magnitude has a wide range in SNR.   

\subsection{Radial velocities}

\begin{figure}
  \centering  
  \includegraphics[width=\columnwidth]{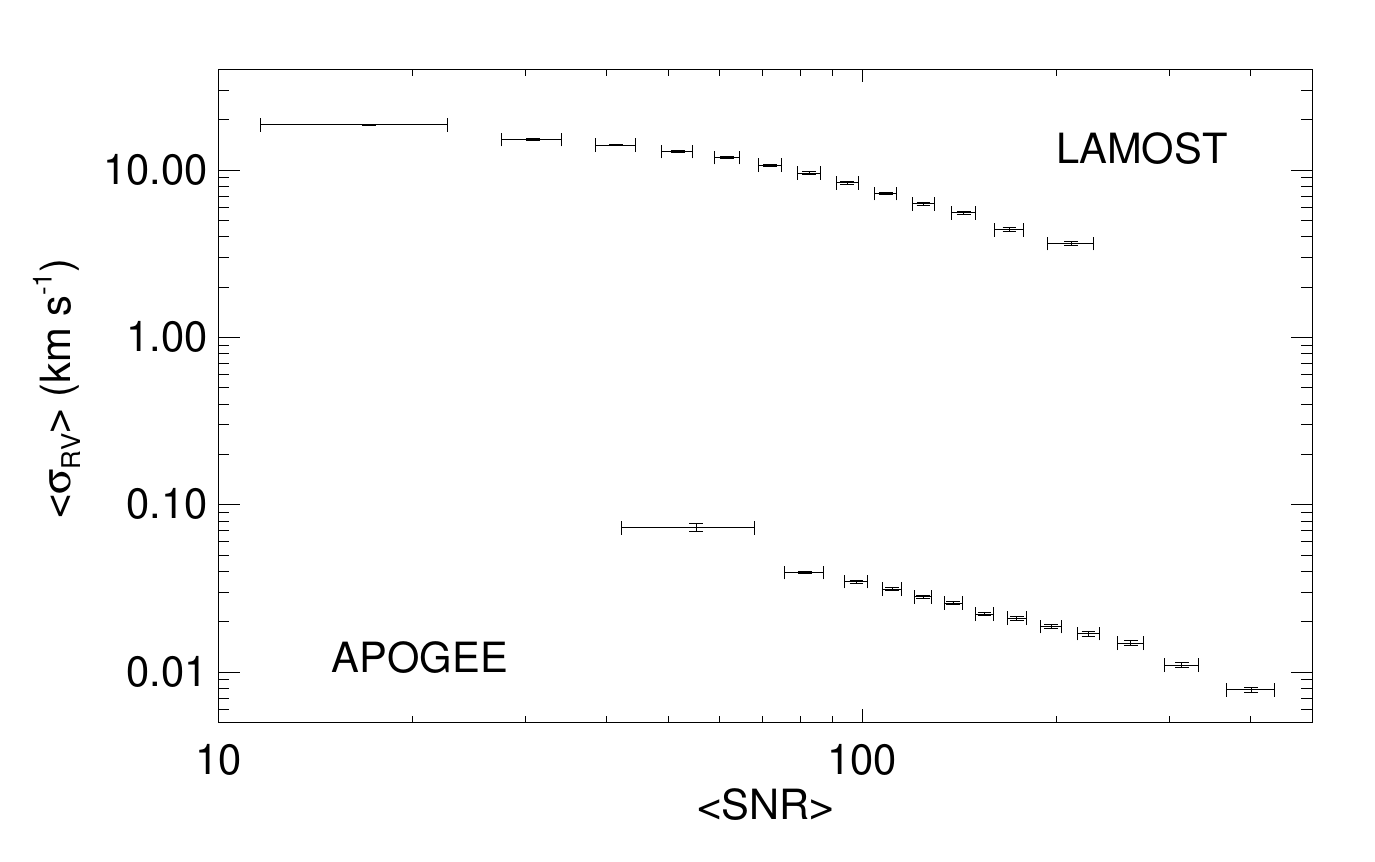}
   \caption{Mean radial velocity uncertainties as a function of the mean SNR for the common stars in APOGEE DR14 and LAMOST DR3. Each point contains 3,000 stars. The error bars represent the standard deviation in SNR and the standard error in $\sigma_{RV}$.}
  \label{fig:erV_SNR}
\end{figure}

\begin{figure}
  \centering  
  \includegraphics[width=\columnwidth]{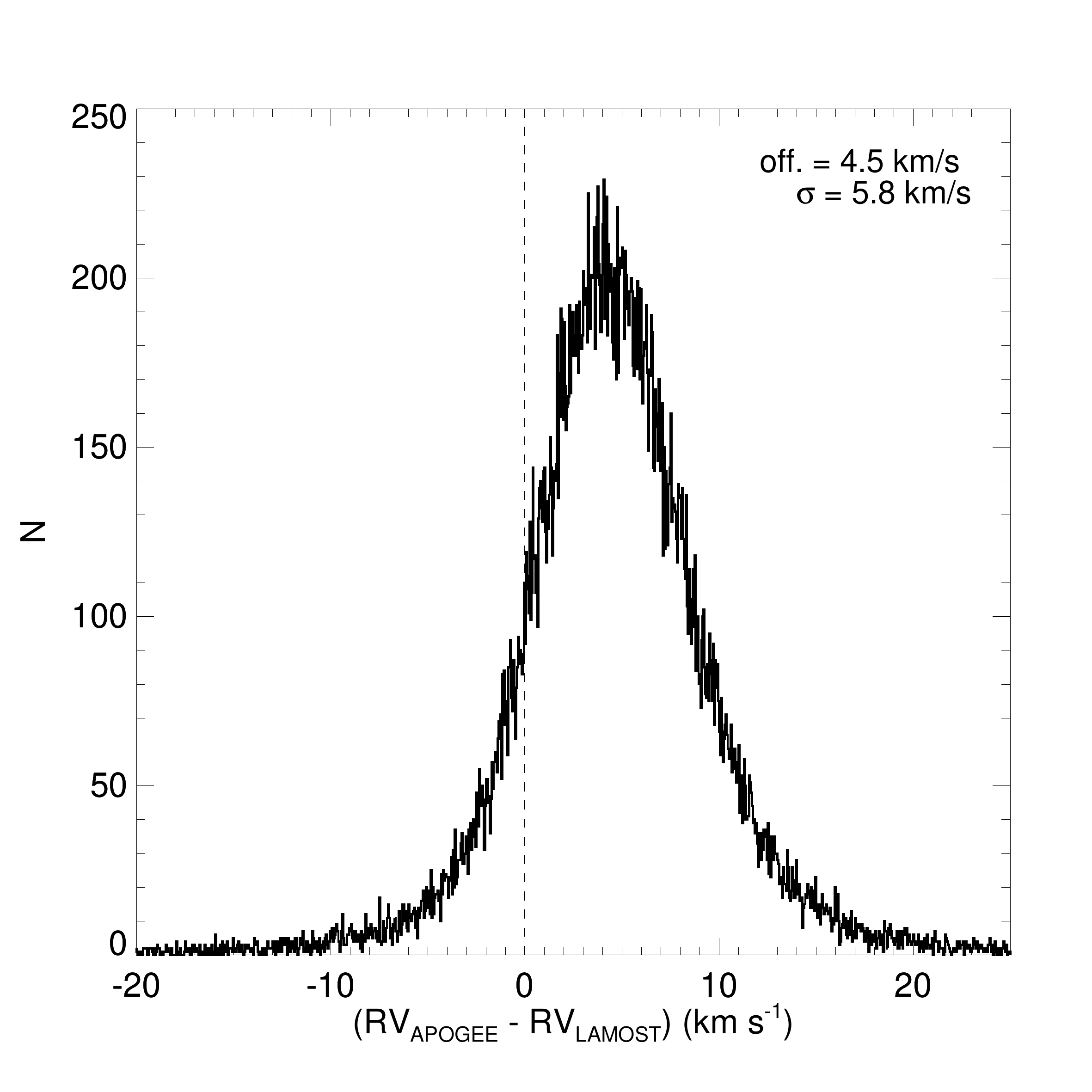}
   \caption{Histograms of discrepancies between APOGEE and LAMOST RVs. The distribution shows a mean offset of about 5 km s$^{-1}$.}
  \label{fig:Histo_RV}
\end{figure}

In APOGEE DR14 \cite{2018ApJS..235...42A} radial velocities (RVs) are determined for each individual visit to identify stars with companion-induced Doppler shifts. The individual visit spectra are resampled and combined to generate a single spectrum for each object. Final RVs are obtained by cross-correlation against a grid of synthetic spectra spanning a wide range of stellar parameters. The APOGEE instrument and the existing radial velocity software routinely deliver radial velocities per visit at a precision of $\sim$ 0.07 km s$^{-1}$ for SNR $>$ 20, while the survey provides external calibration sufficient to ensure accuracies at the level of $\sim$ 0.35 km s$^{-1}$. RVs in APOGEE are reported with respect to the center of mass (barycenter) of the solar system \cite{2015AJ....150..173N}. 

The LAMOST 1D pipeline \cite{2015RAA....15.1095L} also measures the RVs by using a cross-correlation method. The pipeline recognizes the stellar spectral classes and simultaneously determines the RVs from the best fit correlation function between the observed spectra and the template. The RVs are corrected from geocentric coordinates to barycentric coordinates. Figure~\ref{fig:erV_SNR} shows the mean reported RV uncertainties with respect to the mean SNR for the common catalog between APOGEE and LAMOST. Figure~\ref{fig:erV_SNR} shows the standard deviation of the SNR and the error deviation of $\sigma_{RV}$ for bins of 3,000 stars by SNR. Nearly all the stars in the APOGEE catalog in common with LAMOST have $\sigma_{RV}$ $<$ 0.1 km s$^{-1}$. For stars with SNR$_{APOGEE}$ $>$ 100, APOGEE determines $\sigma_{RV}$ $<$ 0.03 km s$^{-1}$. The LAMOST calculated RV uncertainties for stars in common with APOGEE range from 3.5 to 18 km s$^{-1}$, depending on the SNR. Stars with SNR$_{LAMOST}$ $>$ 100, show $\sigma_{RV}$ $<$ 8 km s$^{-1}$ (see Fig.~\ref{fig:erV_SNR}). 

\begin{figure}
  \centering  
  \includegraphics[width=1.\columnwidth]{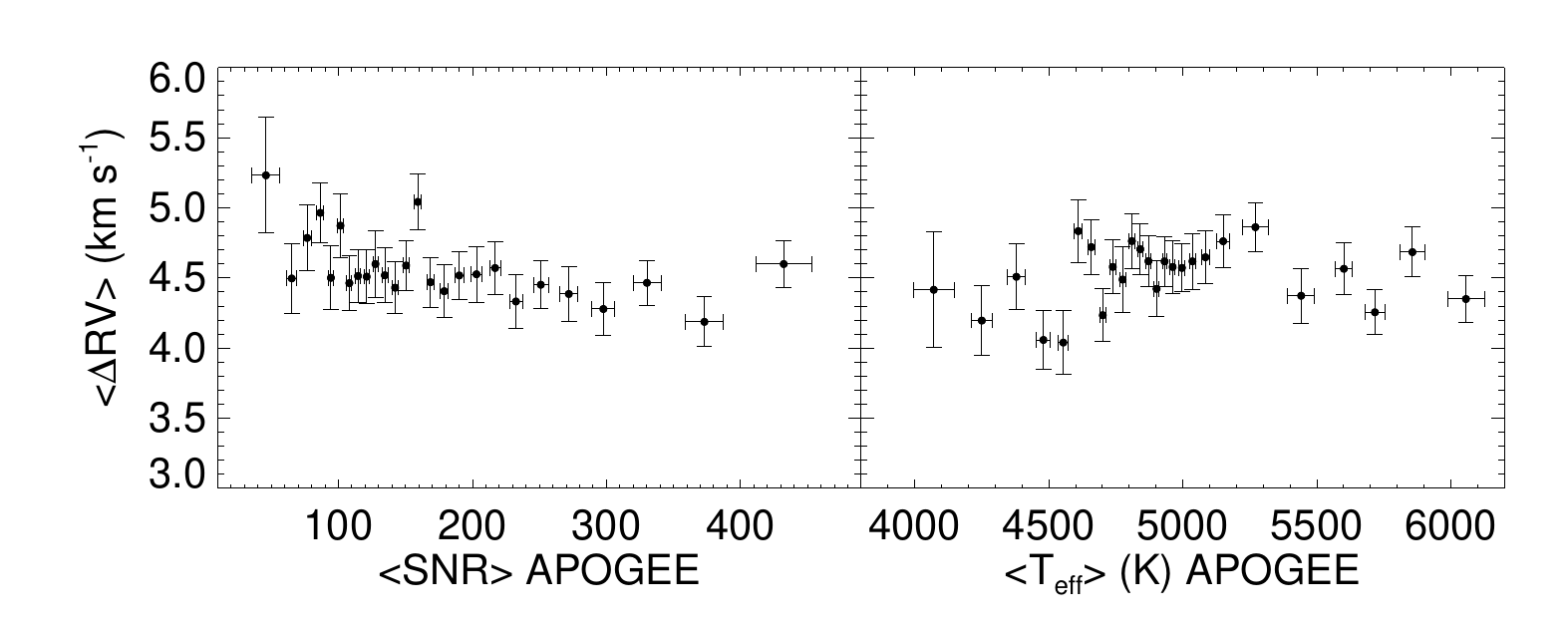}
  \includegraphics[width=1.\columnwidth]{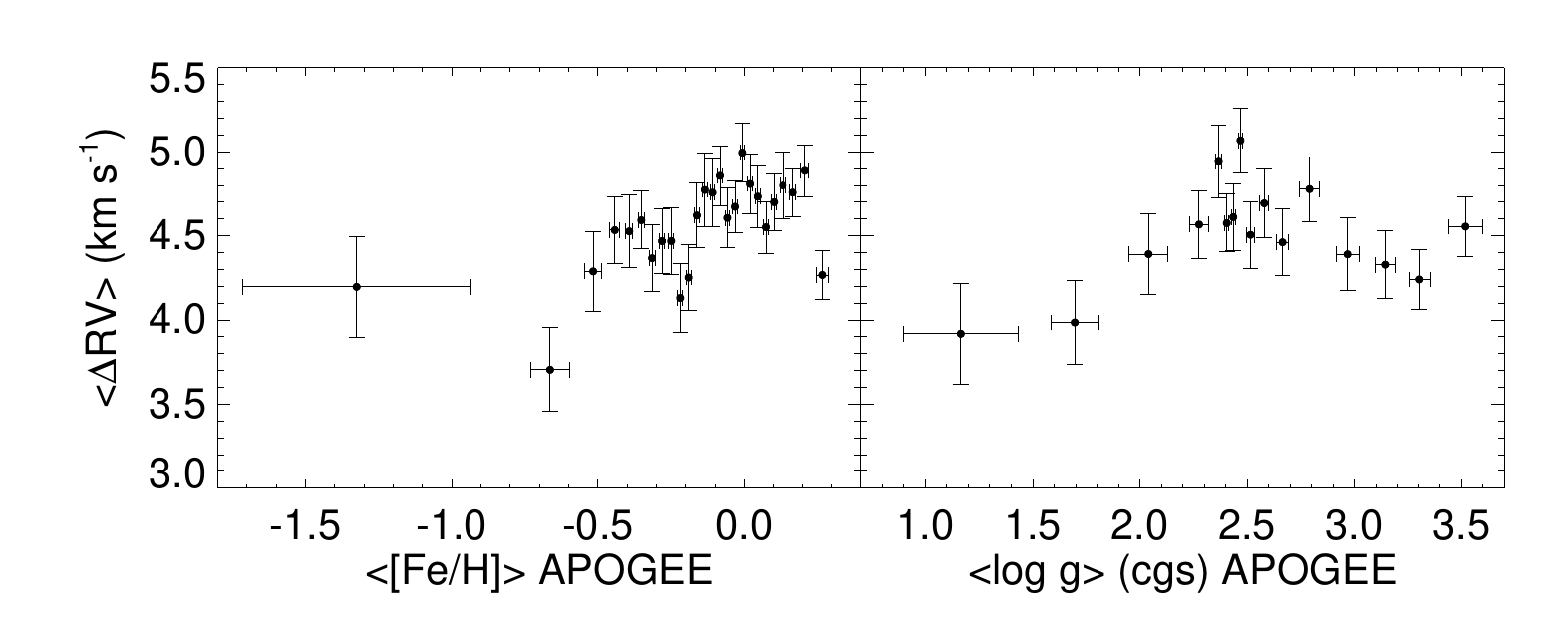}
   \caption{Mean radial velocity discrepancies with respect to $<$SNR$>$, $<T_{eff}>$, $<$[Fe/H]$>$ and $<$log g$>$ from APOGEE DR14. There is no clear systematic trends between the RVs discrepancies and the stellar parameters, where the amplitude is $\leq$ 1 km s$^{-1}$.}
  \label{fig:Mean_RV}
\end{figure}

\begin{figure*}
  \centering  
  \includegraphics[width=2.\columnwidth]{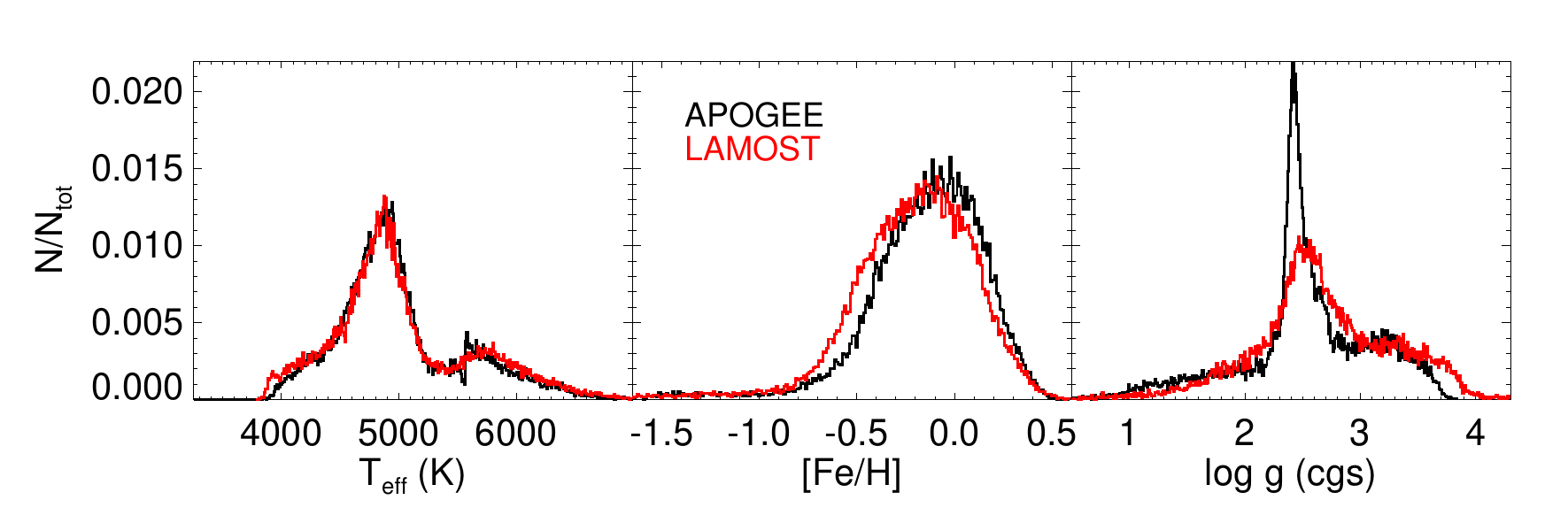}
   \caption{Temperature, metallicity and surface gravity distribution of the APOGEE-LAMOST catalog, using the stellar parameters from APOGEE DR14 (black line), and the parameters reported in LAMOST DR3 (red line).}
  \label{fig:histo_param}
\end{figure*}

\begin{figure}
  \centering  
  \includegraphics[width=\columnwidth]{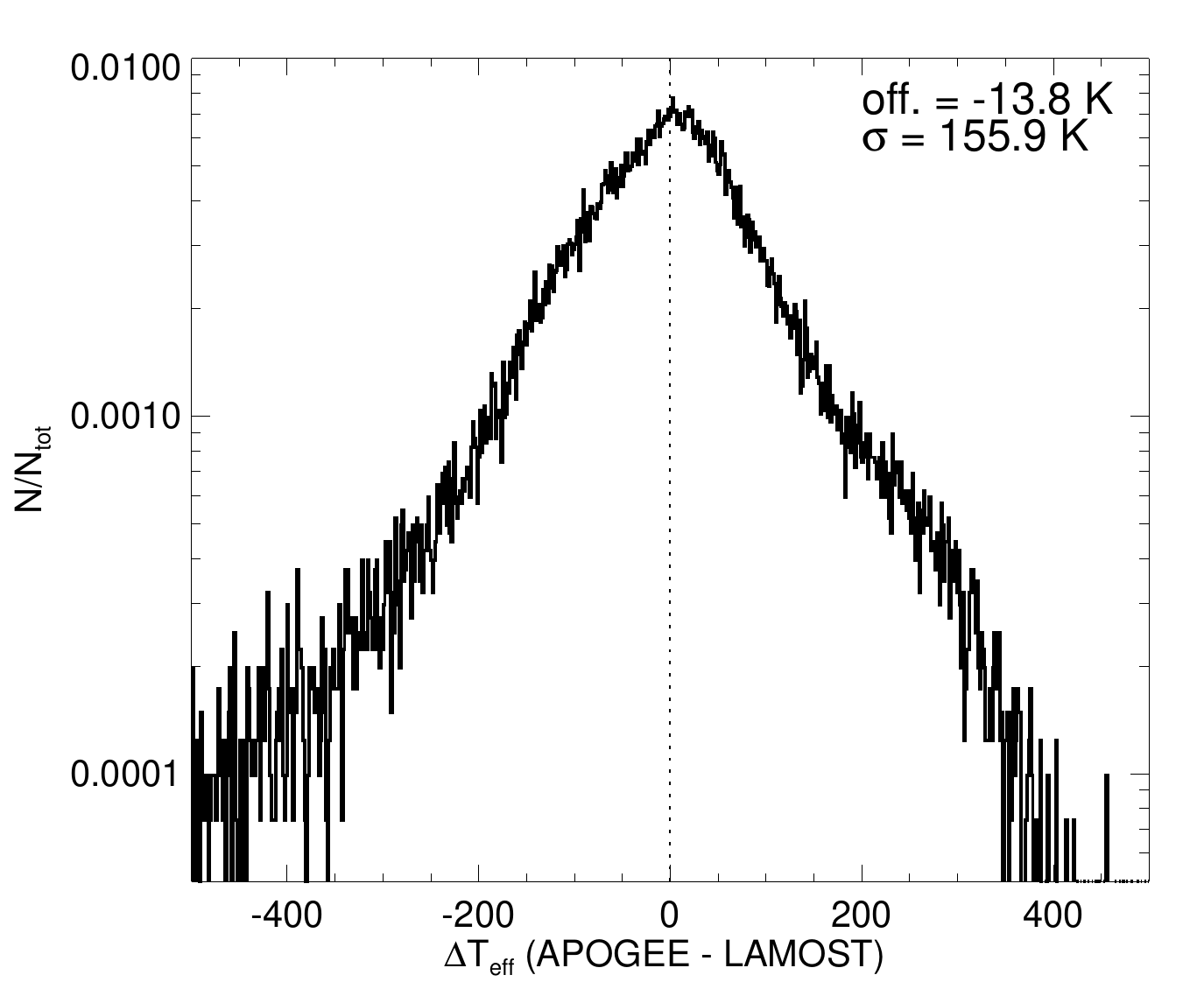}
   \includegraphics[width=\columnwidth]{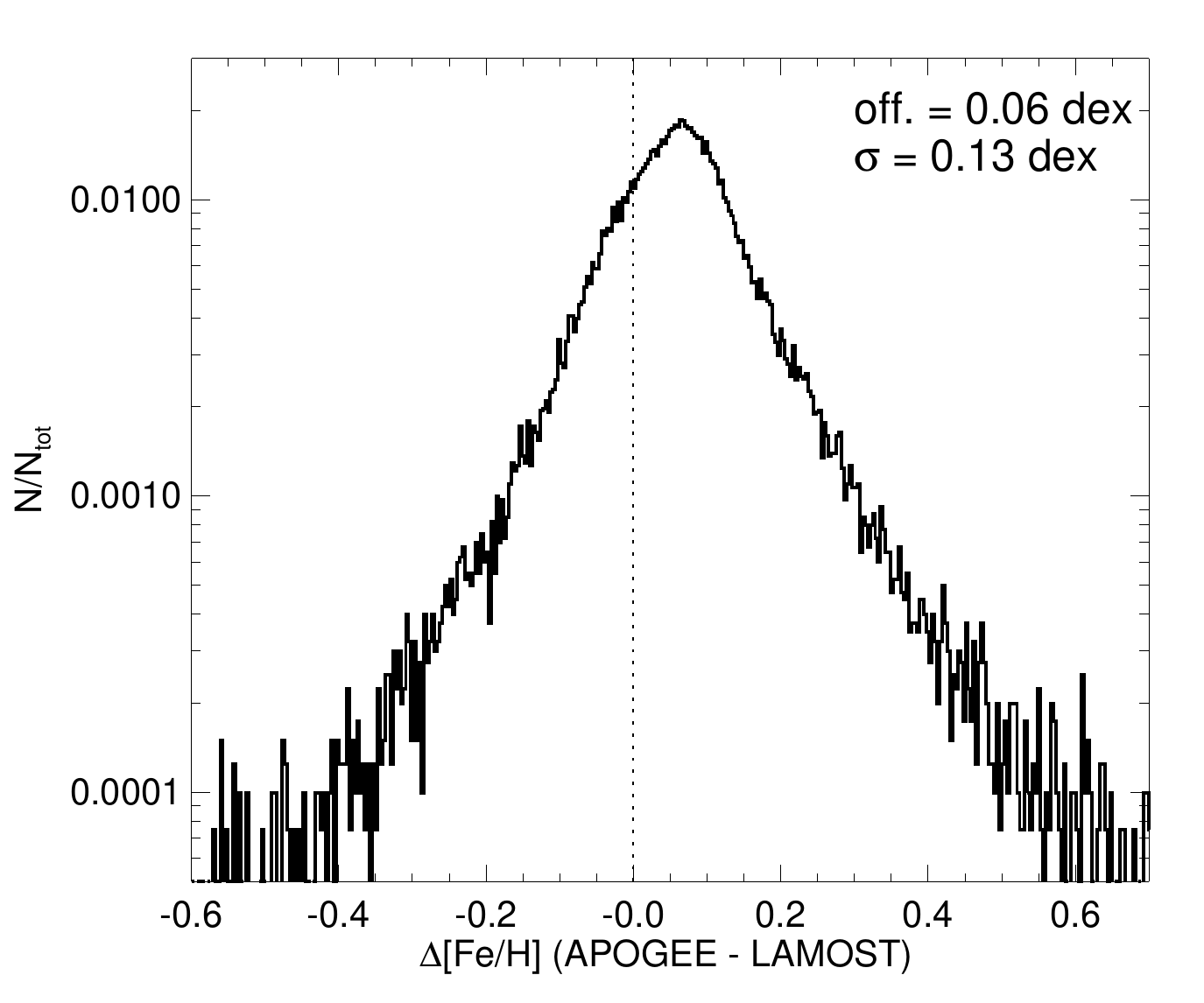}
   \includegraphics[width=\columnwidth]{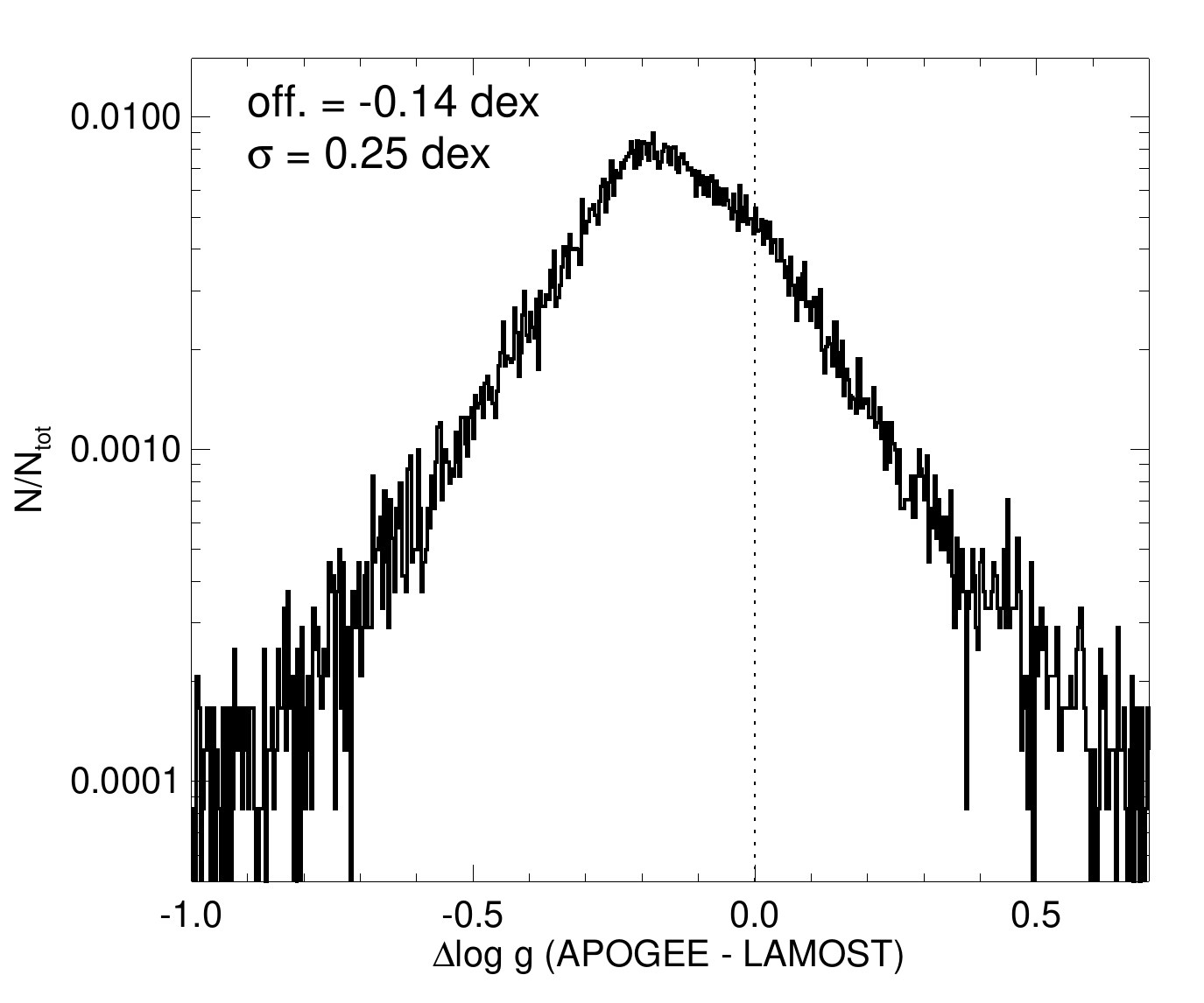} 
   \caption{Histograms of discrepancies between effective temperature (upper panel), metallicity (middle panel) and surface gravity (bottom panel) in logarithmic scale, respectively. There is a clear offset between APOGEE and LAMOST for the surface gravity, also a substantial scatter of 0.25 dex. See text for details.}
  \label{fig:histo_discre}
\end{figure}

In Figure~\ref{fig:Histo_RV} we have the histogram of discrepancies between APOGEE and LAMOST RVs. The discrepancies show a clear offset of 4.54 $\pm$ 0.03 km s$^{-1}$, with a dispersion of 5.8 km s$^{-1}$. Most of the scatter in the discrepancies is expected to come from LAMOST RVs uncertainties (Fig.~\ref{fig:erV_SNR}), which suggests that the average LAMOST measurement error for the RVs is $\sim$ 6 km s$^{-1}$. This is consistent with the median in the reported RV LAMOST uncertainty, where the value is $\sim$ 6.5 km s$^{-1}$. \cite{2015MNRAS.449..162H} compared 499 RVs from LAMOST DR1 with RVs of the same targets derived from MMT+Hectospec, and also found an offset, in their case of 3.8 $\pm$ 0.3 km s$^{-1}$. Furthermore, a similar offset appears between an external comparison between LAMOST DR1 and SEGUE DR9 for common objects, where the result is $<\Delta$RV$>$ = 7.2 km s$^{-1}$.  A similar offset is also reported in \cite{2017MNRAS.472.3979S} for the LAMOST survey. 


In Figure~\ref{fig:Mean_RV} we explore the mean RVs discrepancies as a function of $<$SNR$>$, $<T_{\rm eff}>$, $<$[Fe/H]$>$ and $<$log g$>$ from APOGEE DR14. We find no clear systematic trends between the RVs discrepancies and the stellar parameters, except for a weak trend of amplitude 1 km s$^{-1}$ as a function of $<$[Fe/H]$>$ (lower-left panel in Fig.~\ref{fig:Mean_RV}). From the lack of trends in what we are able to compare between the datasets, the origin of the global RV offset between the surveys is unclear.

\subsection{Stellar atmospheric parameters}
\label{sp}

Stellar parameters are derived from the combined APOGEE spectra with the APOGEE Stellar Parameters and Chemical Abundances Pipeline (ASPCAP) (\cite{2016AJ....151..144G}), where an interpolated grid of synthetic spectra (e.g., \cite{2015AJ....149..181Z}) is searched to find the best match to each observed spectrum. ASPCAP performs a a multidimensional $\chi^{2}$ minimization using the code FERRE\footnote{github.com/callendeprieto/ferre}. We emphasize that for this comparison we use APOGEE \emph{calibrated} stellar parameters. ASPCAP derive parameters for nearly all the observed stars, but these parameters suffer for systematic errors, most likely associated to shortcomings in the models. Hence, APOGEE produce \emph{calibrated} parameters, i.e. surface gravities calibrated to seismic gravities via asteroseismology for giant stars. We refer the reader to \cite{2015AJ....150..148H} and \cite{2016AJ....151..144G}, where ASPCAP and calibrated parameters are explained in detail.   

The LAMOST Stellar Parameter Pipeline (LASP) also determines automatically the fundamental stellar parameters from LAMOST spectra. LASP adopts two methods to obtain a single set of derived stellar parameters: (i) the Correlation Function Initial (CFI) values \cite{2012SPIE.8451E..37D} followed by the ULySS package \cite{2011RAA....11..924W} and (ii) the CFI method is used to get an initial guess for the stellar parameters, while ULySS generates the final values from the observed spectra \cite{2015RAA....15.1095L}. 

In Figure~\ref{fig:histo_param} we have the stellar atmospheric parameters for temperature, metallicity and surface gravity reported in APOGEE DR14 (black line), and in LAMOST DR3 (red line) for the APOGEE - LAMOST catalog. The histograms show the stellar parameters range we compare in this exercise. Figure~\ref{fig:histo_discre} shows the histograms of discrepancies between derived T$_{\rm eff}$, [Fe/H] and log g in logarithmic scale from the APOGEE and LAMOST surveys for the objects in common. We observe a small offset in effective temperature of about 13 K, with a scatter of 155 K. We also observe a small offset in [Fe/H] of about 0.06 dex together with a scatter of 0.13 dex. Using only surface gravities in calibrated red giants from APOGEE DR14, where there are 24,074 stars in common, we notice that the largest offset between both surveys occurs in the surface gravities, where a deviation of 0.14 dex is observed with a substantial scatter of 0.25 dex. Holtzman et al. (2018, in prep.) reported that at low surface gravities the APOGEE ASPCAP spectroscopy surface gravities are systematically higher than the asteroseismic surface gravities; as a result the log g in APOGEE DR14 is calibrated using a linear surface gravity fit to the seismic log g measured for giants in the first APOKASC catalog \cite{2014ApJS..215...19P}. This calibration could explain the offset reported here with respect to the LAMOST surface gravities. More details about the spectroscopic and photometric NASA \emph{Kepler} field follow-ups is given in Section~\ref{K2}. Figure~\ref{fig:histo_discre} also shows that the histograms of discrepancies between the measured stellar parameters in APOGEE and LAMOST are not symmetric, which suggests that other systematic effects may be present in the data. 

\begin{figure}
  \centering  
  \includegraphics[width=\columnwidth]{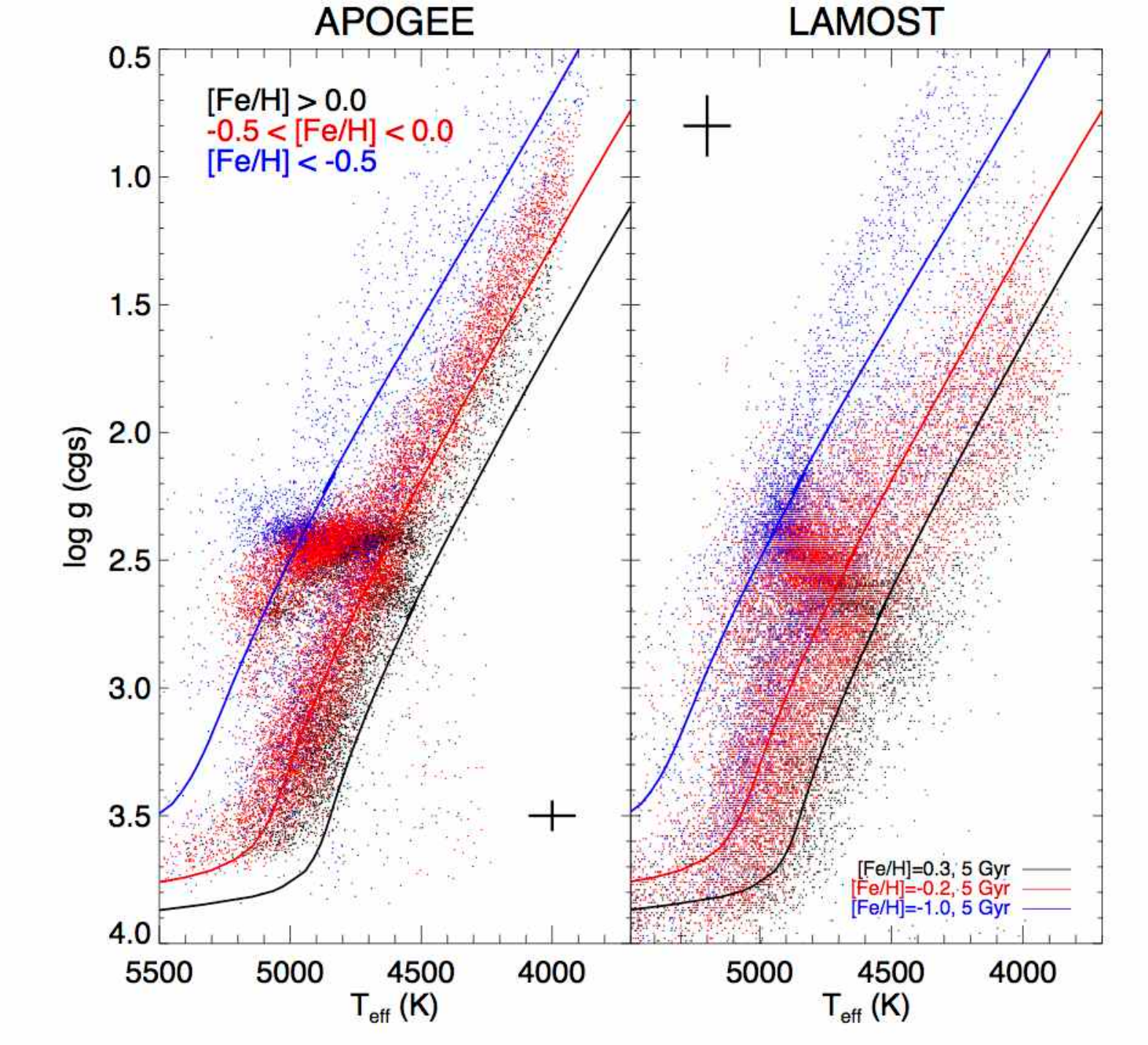}
   \caption{T$_{\rm eff}$ - log g diagram from APOGEE DR14 (left) and LAMOST DR3 (right) stellar atmosphere parameters for the stars in common, color coded by three different ranges in [Fe/H]. Over plotted are 5 Gyr isochrones \cite{2012MNRAS.427..127B} at [Fe/H] = +0.3, --0.2 and --1.0 dex, respectively. The errorbars represent the typical uncertainties in T$_{\rm eff}$ and $\log$ g in the two surveys.}
  \label{fig:Kiel}
\end{figure}

Figure~\ref{fig:Kiel} shows the T$_{\rm eff}$ - log g diagram from the APOGEE DR14 and LAMOST DR3 stellar parameters, color coded by [Fe/H]. Over-plotted are 5 Gyr isochrones \cite{2012MNRAS.427..127B} at three different metallicities as indicated in the figure. There is a generally good agreement between the theoretical stellar tracks and the data-sets from both surveys. However, the Red Clump (RC) appears clearly distint in the APOGEE T$_{\rm eff}$ - log g diagram around log g $\sim$ 2.5, while the clump in the LAMOST data is more diffuse, showing a spread from 2.2 to 2.7 dex in surface gravity. In the next sections we explore the reported uncertainties associated with the atmosphere stellar parameters.

\begin{figure}
  \centering  
  \includegraphics[width=\columnwidth]{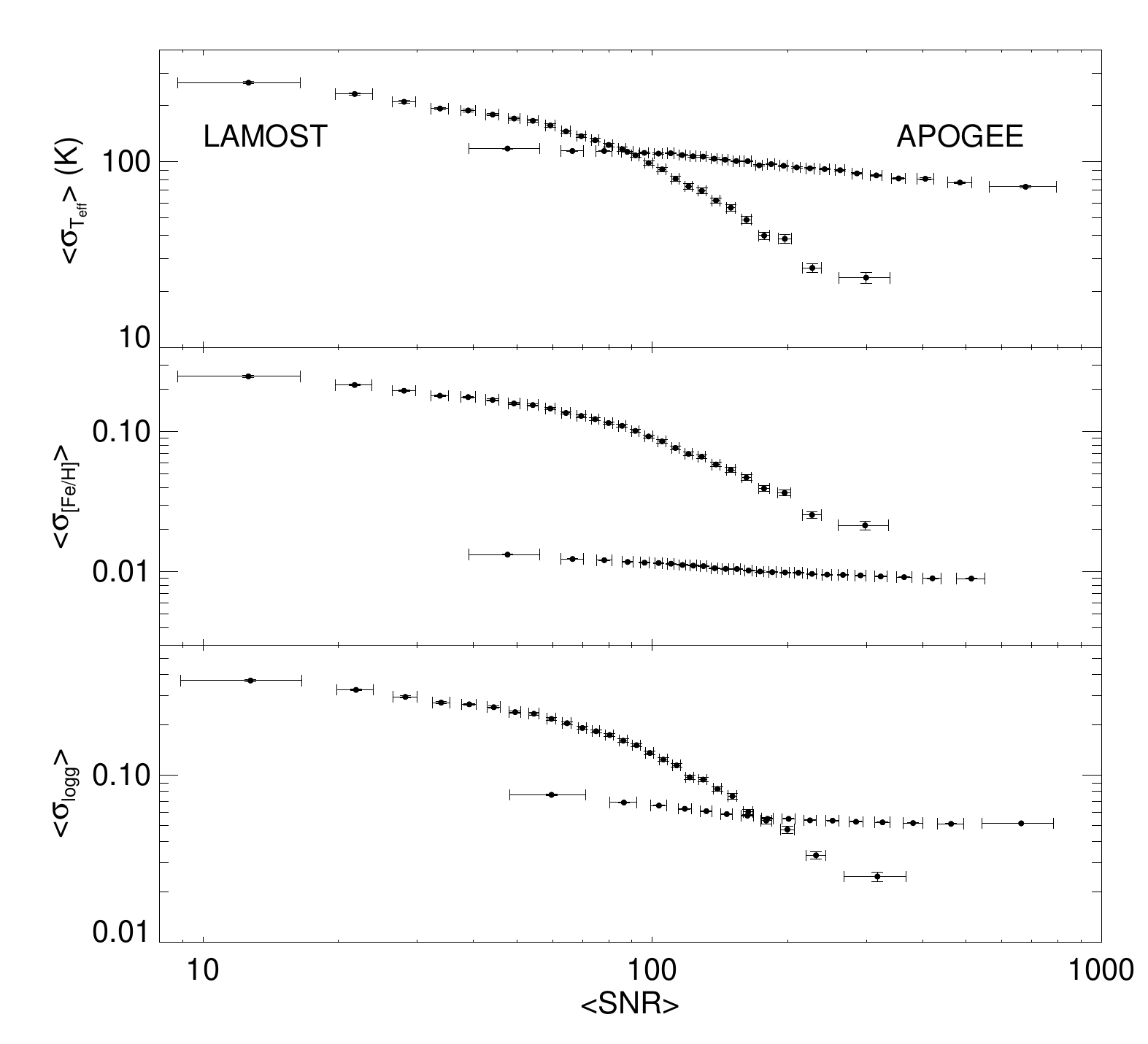}
   \caption{Average uncertainty in T$_{\rm eff}$, [Fe/H] and log g as a function of the average APOGEE $<SNR>$ for stars in common in the APOGEE and LAMOST survey. Every point has 1500 stars and the error bars represent the standard deviation in the SNR bin and the standard error of the mean for T$_{\rm eff}$, [Fe/H] and log g, respectively.}
  \label{fig:uncert_para}
\end{figure}

\subsubsection{Uncertainties}
\label{uncer}

In Figure~\ref{fig:uncert_para} we explore the quoted uncertainties in the main atmosphere stellar parameters in the APOGEE-LAMOST stellar catalog. The three panels show the reported uncertainties in T$_{\rm eff}$, [Fe/H] and log g as a function of the SNR for both surveys. 

\begin{table*}[htp]
\caption{Mean offset from the histogram of discrepancies, the observed scatter and the sum of uncertainties in both catalogs for different SNR range in the LAMOST survey.}
\begin{center}
\begin{tabular}{ccccccccccc}
\hline
SNR$_{\rm L}$ & $<\Delta T_{\rm eff}>$ & $\sigma_{T_{\rm eff}}$ & $(\sigma_{\rm A}^{2} + \sigma_{\rm L}^{2})^{1/2}$ & $<\Delta[Fe/H]>$ & $\sigma_{\rm [Fe/H]}$ & $(\sigma_{\rm A}^{2} + \sigma_{\rm L}^{2})^{1/2}$ & $<\Delta log g>$ & $\sigma_{\rm log g}$ & $(\sigma_{\rm A}^{2} + \sigma_{\rm L}^{2})^{1/2}$ \\
\hline
\hline

$<$ 100 &  --2.92 &  146.82  &  191.58 &  0.07 &  0.13 &  0.15 & --0.13 &  0.25  &  0.25 \\
$>$ 200 &  --30.09 &     119.47  &    97.28  & 0.05  &  0.09 &  0.03 & --0.17 &  0.21 & 0.07\\

\end{tabular}
\end{center}
\label{tab:uncert}
\end{table*}%

\begin{figure}
  \centering  
  \includegraphics[width=1.\columnwidth]{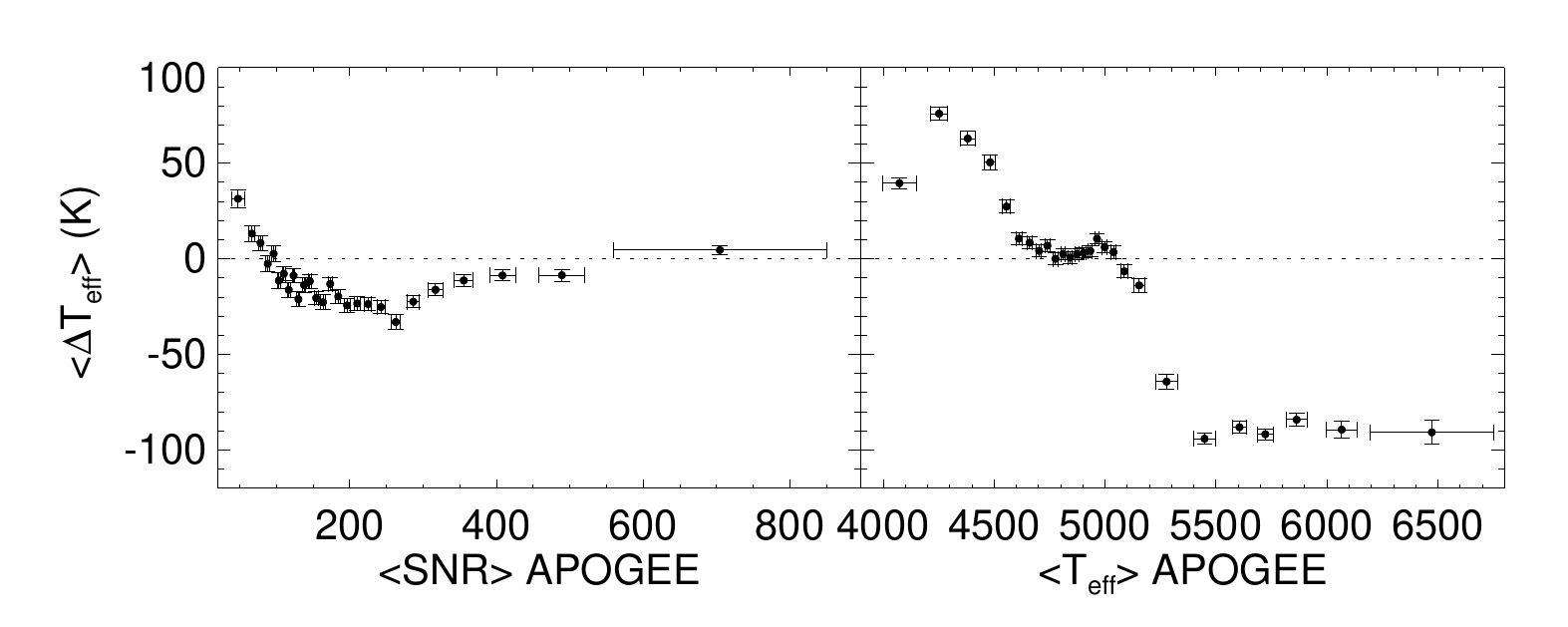}
   \includegraphics[width=1.\columnwidth]{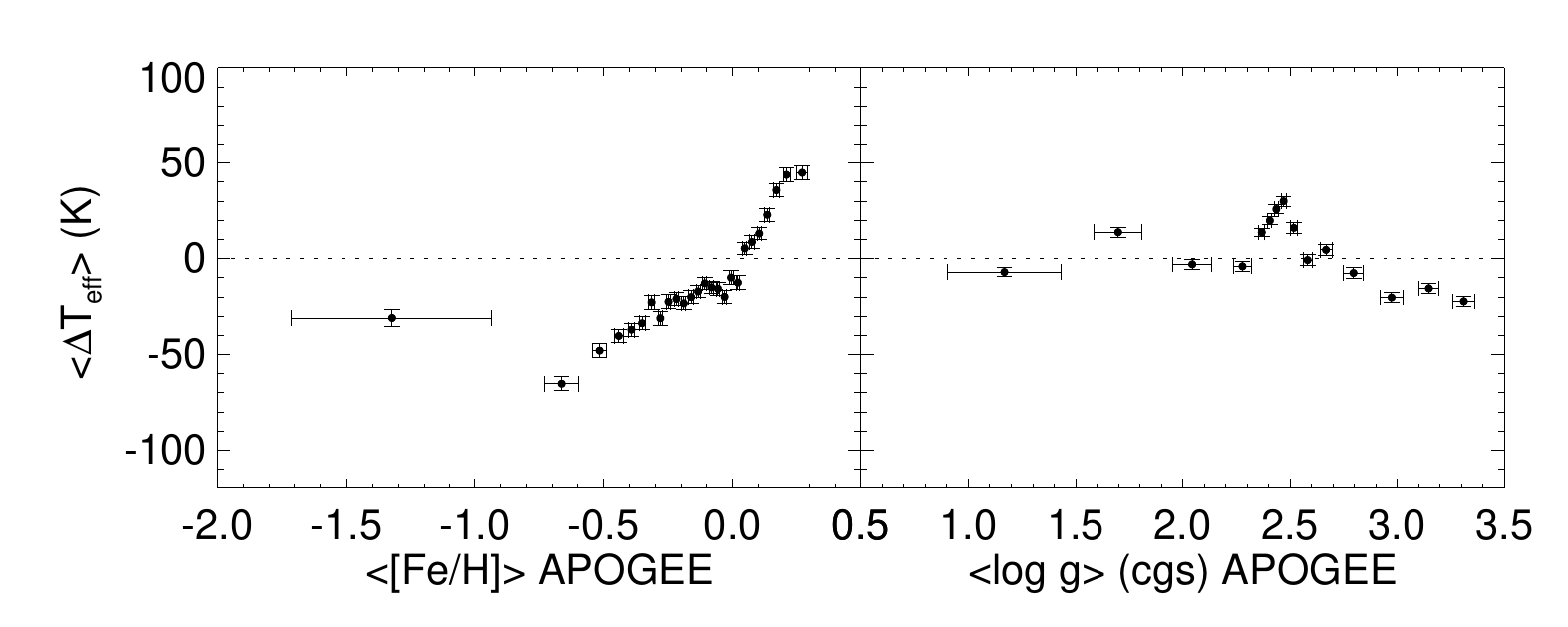}
   \caption{Mean effective temperature discrepancies (APOGEE - LAMOST) as a function of $<$SNR$>$, $<T_{\rm eff}>$, $<$[Fe/H]$>$ and $<$log g$>$.}
  \label{fig:T_sys}
\end{figure}

\begin{figure}
  \centering  
  \includegraphics[width=1.\columnwidth]{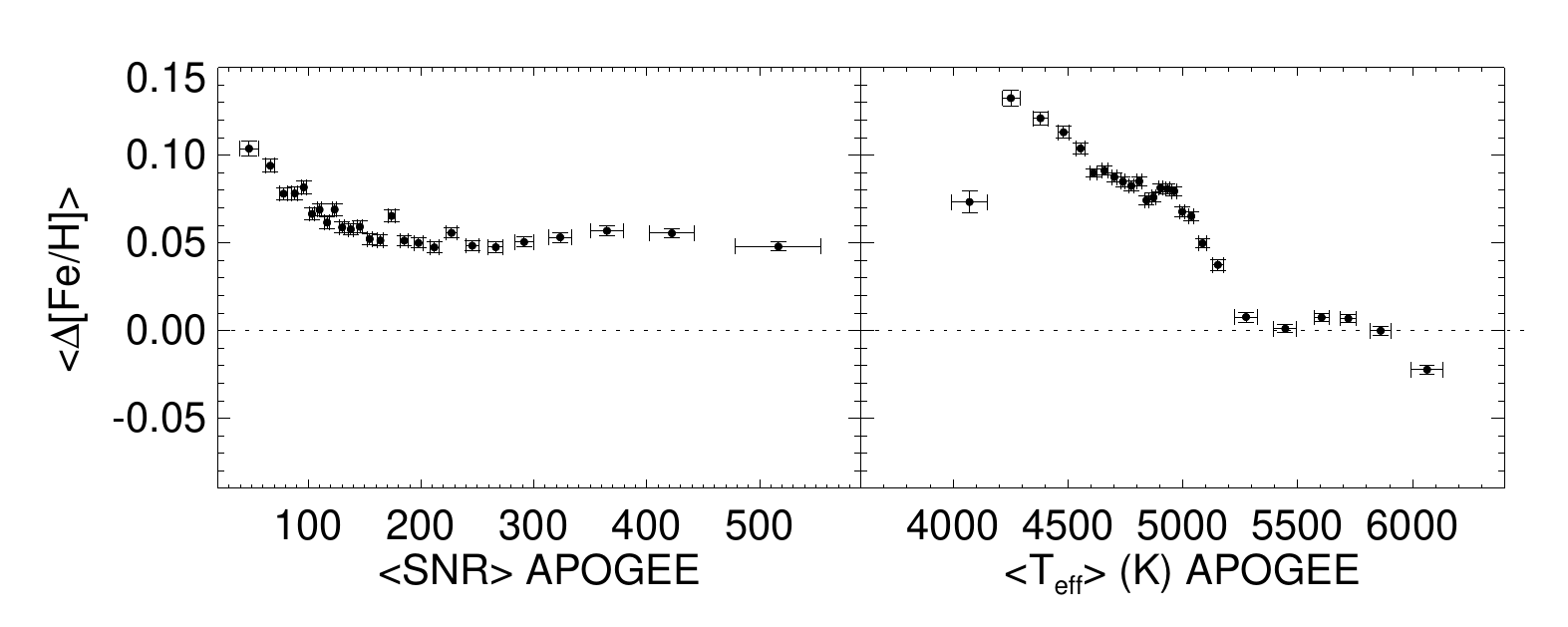}
   \includegraphics[width=1.\columnwidth]{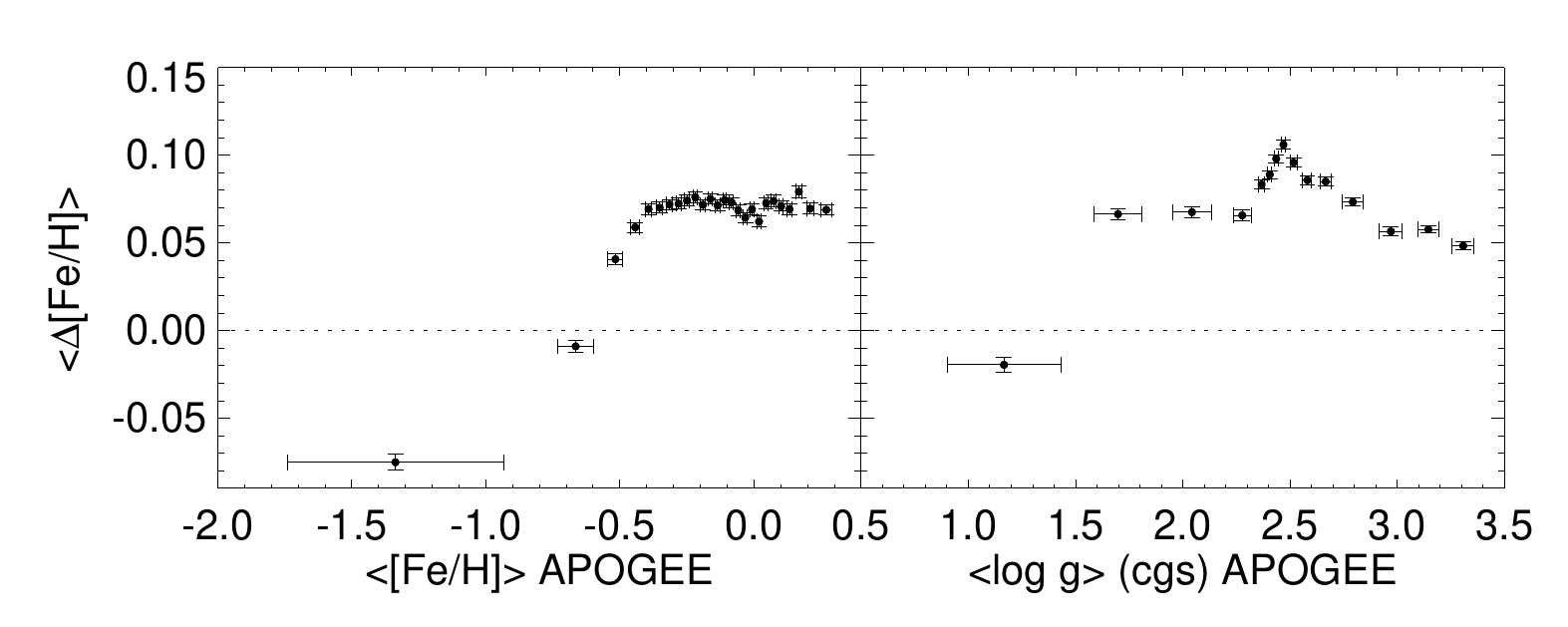}
   \caption{Mean [Fe/H] discrepancies (APOGEE - LAMOST) as a function of $<$SNR$>$, $<T_{\rm eff}>$, $<$[Fe/H]$>$ and $<$log g$>$.}
  \label{fig:Fe_sys}
\end{figure}

\begin{figure}
  \centering  
  \includegraphics[width=1.\columnwidth]{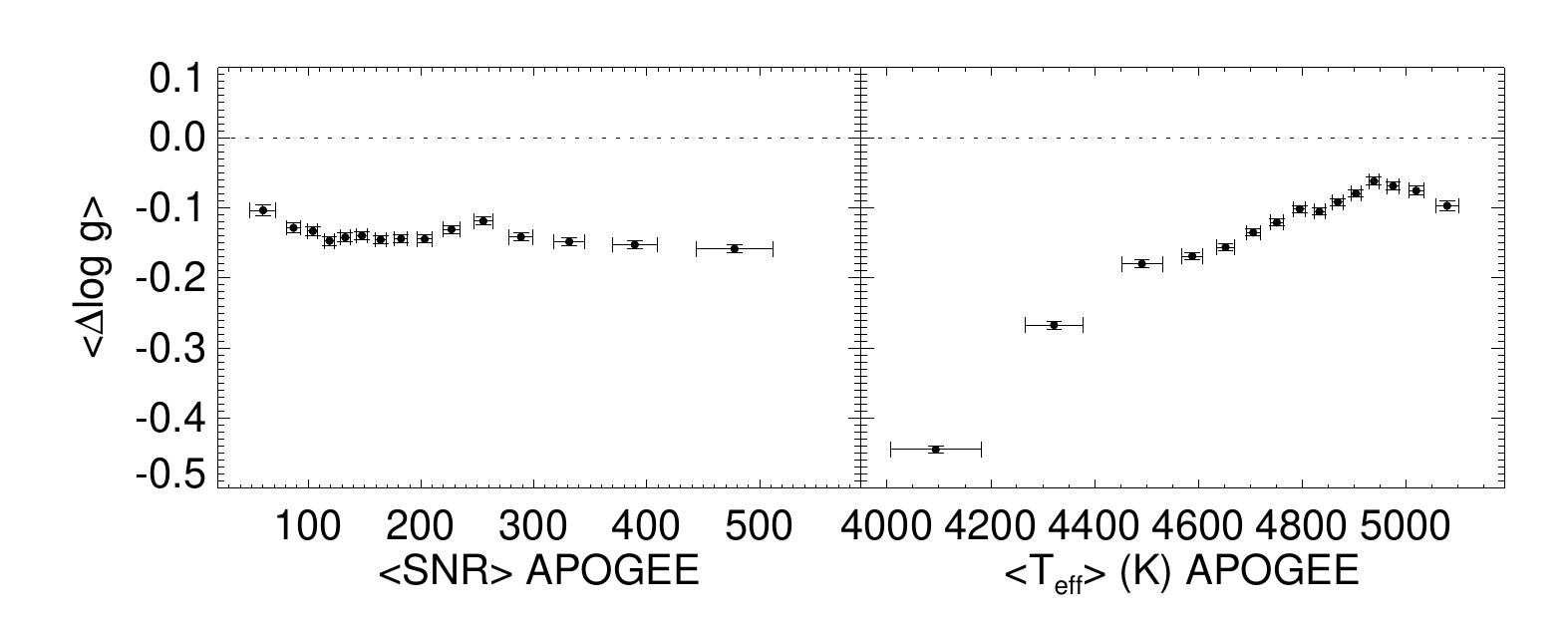}
   \includegraphics[width=1.\columnwidth]{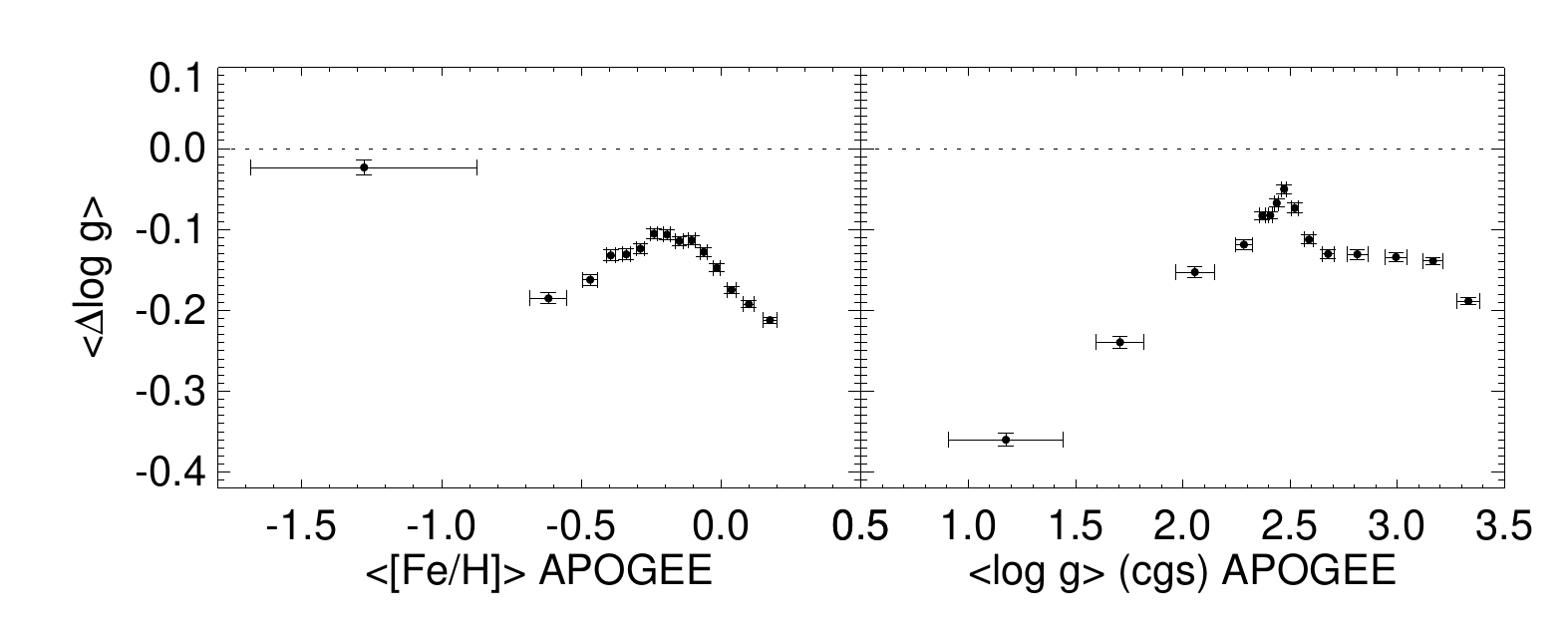}
   \caption{Mean surface gravity discrepancies (APOGEE - LAMOST) with respect to $<$SNR$>$, $<T_{\rm eff}>$, $<$[Fe/H]$>$ and $<$log g$>$, respectively.}
  \label{fig:grav_sys}
\end{figure}

From the entire sample in common we calculated the mode of the reported uncertainty distribution for the three atmospheric parameters. The median APOGEE uncertainties are $\sim$ 90 K, $\sim$ 0.01 dex and $\sim$ 0.06 dex in T$_{\rm eff}$, [Fe/H] and log g, respectively. For LAMOST, we find $\sim$ 80 K, $\sim$ 0.08 dex and also $\sim$ 0.12 dex, respectively. From these later values, the reported LAMOST DR3 uncertainties seem underestimated. Interestingly, the uncertainties in the APOGEE DR14 stellar parameters are nearly independent of the SNR, asymptoting to a ``floor" due to systematics. On the other hand, the uncertainties reported in the LAMOST DR3 show a clear dependence on the SNR, with no clear ``floor" reached at the highest SNR. For the effective temperature, stars with SNR$_{LAMOST}$ $>$ 100 show that, on average, $\sigma_{\rm Teff_{\rm lamost}}$ $<$ $\sigma_{\rm Teff_{\rm apogee}}$. For iron abundances, we have that APOGEE uncertainties are, on average, lower than the iron uncertainties reported in LAMOST independent of the SNR of the spectrum, while stars with SNR$_{LAMOST}$ $>$ 150 report on average, more precise surface gravity uncertainties than those claimed for the APOGEE sample (Fig.~\ref{fig:uncert_para}). We use the histograms of discrepancies between stellar parameter to understand the true error values behind the data, where we assume that the observed scatter constrains the stellar parameter uncertainties. In Table~\ref{tab:uncert} we show the mean offset determined from the histograms of discrepancies, the observed scatter and the sum of the uncertainties in both catalogs. Although, the discrepancies are not perfect gaussians due to systematic effects in the data (see Fig.~\ref{fig:histo_discre}), we assume that $\sigma \sim (\sigma_{A}^{2} + \sigma_{L}^{2})^{1/2}$. In Table~\ref{tab:uncert} we have that for stars with SNR$_{\rm LAMOST}$ $>$ 200, the intrinsic scatter from the discrepancies is always larger than ($\sigma_{A}^{2} + \sigma_{L}^{2})^{1/2}$. Moreover, we saw in Figure~\ref{fig:uncert_para} that APOGEE stellar parameters are nearly independent of SNR. Hence we conclude that at high SNR ($>$100) the uncertainties from LAMOST are underestimated. 
 
\subsubsection{Systematic effects}	

We use the APOGEE-LAMOST stellar catalog to explore the existence of systematic errors in the atmosphere stellar parameters derived using the respective survey stellar pipelines described above. 

In Figure~\ref{fig:T_sys} we show the discrepancies in T$_{\rm eff}$ between the surveys for stars in common as a function of $<$SNR$>$, $<T_{\rm eff}>$, $<$[Fe/H]$>$ and $<$log g$>$, where every point bins 1500 stars. $\Delta T_{\rm eff}$ has small variations with respect to the APOGEE SNR, with an amplitude smaller than 80 K. Interestingly, we observe that $\Delta T_{\rm eff}$ changes sign for stars colder than $\sim$ 4500 K and for stars hotter than $\sim$ 5200 K. The offset for the cold stars is around $\sim$ 50 K, however for hot stars it reaches values of 100 K (see top panels in Fig.~\ref{fig:T_sys}). There is a clear trend between $\Delta T_{\rm eff}$ and $<$[Fe/H]$>$, where the sign of $\Delta T_{\rm eff}$ flips at solar metallicity. There is no evident trend between $<$log g$>$ and $\Delta T_{\rm eff}$ (see lower panels in Fig.~\ref{fig:T_sys}). 

In Figure~\ref{fig:Fe_sys} we see how the average discrepancy in [Fe/H] changes with respect to the $<$SNR$>$ and the main stellar parameters. There is an evident offset of $\sim$ 0.06 dex already reported above. This offset is nearly independent of the $<$SNR$>$ in the APOGEE spectra, although for stars where the SNR $<$ 100 the offset goes from 0.06 to 0.1 dex (see top-left panel in Fig.~\ref{fig:Fe_sys}). We also find a clear relation between $<T_{\rm eff}>$ and $\Delta$[Fe/H]. The metallicity offset is larger for colder stars, while for stars with T$_{\rm eff}$ $>$ 5200 K, $\Delta$[Fe/H] $\sim$ 0.0 dex (top-right panel in Fig.~\ref{fig:Fe_sys}). $\Delta$[Fe/H] is independent of the $<$[Fe/H]$>$ for the range --0.5 $<$ [Fe/H] $<$ 0.5 dex, however there is a relation between the two quantities for stars more metal-poor than --0.5 dex. Interestingly, for stars in the surface gravity range 1.0 $<$ log g $<$ 1.5 we do not see the offset in [Fe/H] (lower panels in Fig.~\ref{fig:Fe_sys}). 

Finally, Figure~\ref{fig:grav_sys} shows $\Delta$log g as a function of the main stellar parameters from APOGEE and the average SNR. As in the previous figures each point represents 1500 stars and the errors bars are the standard error of the mean for log g and the standard deviation in the SNR. As we mentioned above there is an offset of $\sim$ 0.15 dex in $\Delta$log g. This offset is independent of the SNR of the observed spectra. For stars colder than 4800 K we observe a dependence between $\Delta$log g and T$_{\rm eff}$. In the lower panels of Figure~\ref{fig:grav_sys}, $\Delta$log g as a function of $<$[Fe/H]$>$ shows variations of about 0.1 dex. We also find a dependence between $<$log g$>$ and $\Delta$log g for stars with log g $<$ 2.5 dex. 




\section{The NASA \emph{Kepler} field}
\label{K2}

\begin{figure}
  \centering  
  \includegraphics[width=\columnwidth]{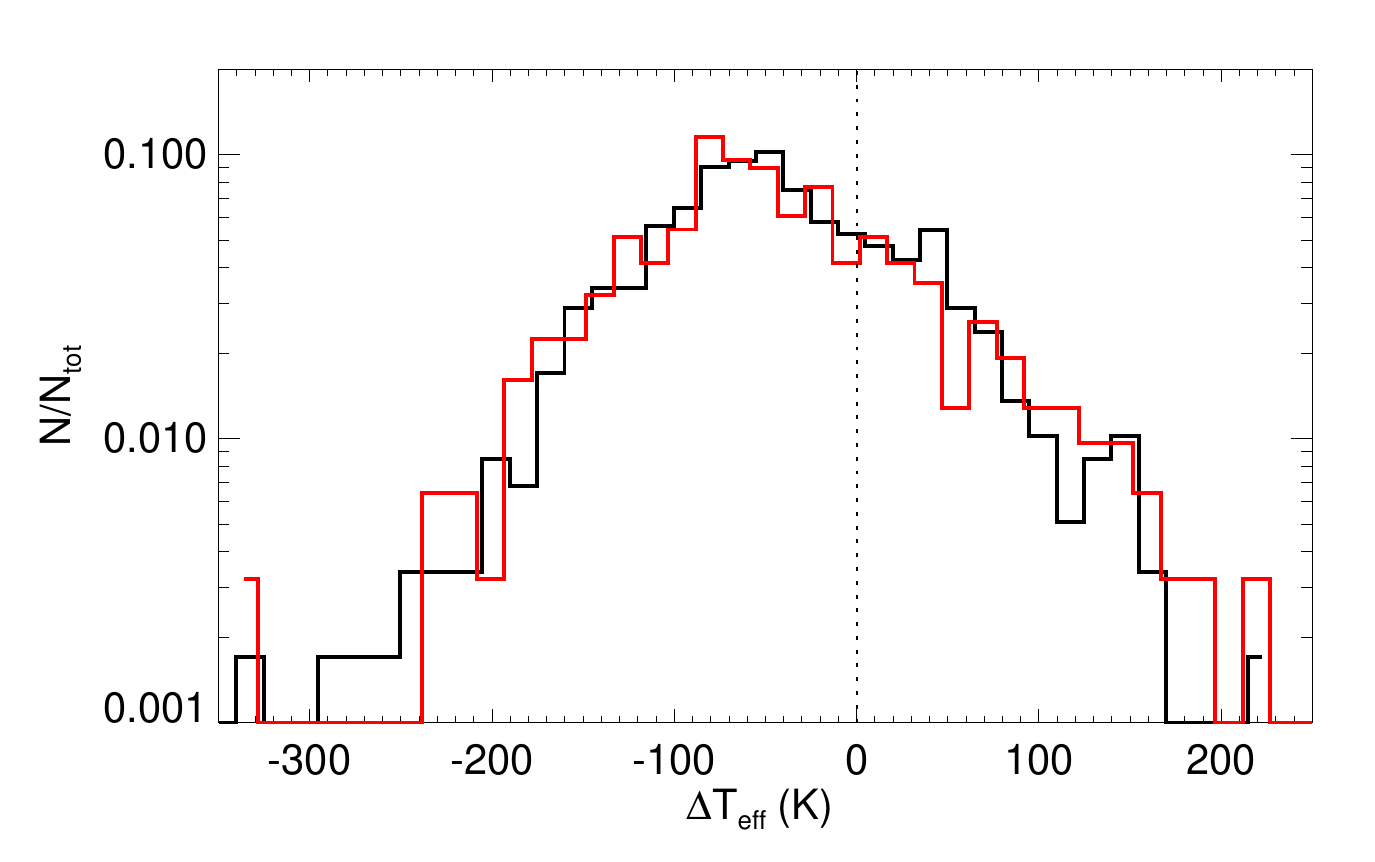}
   \includegraphics[width=\columnwidth]{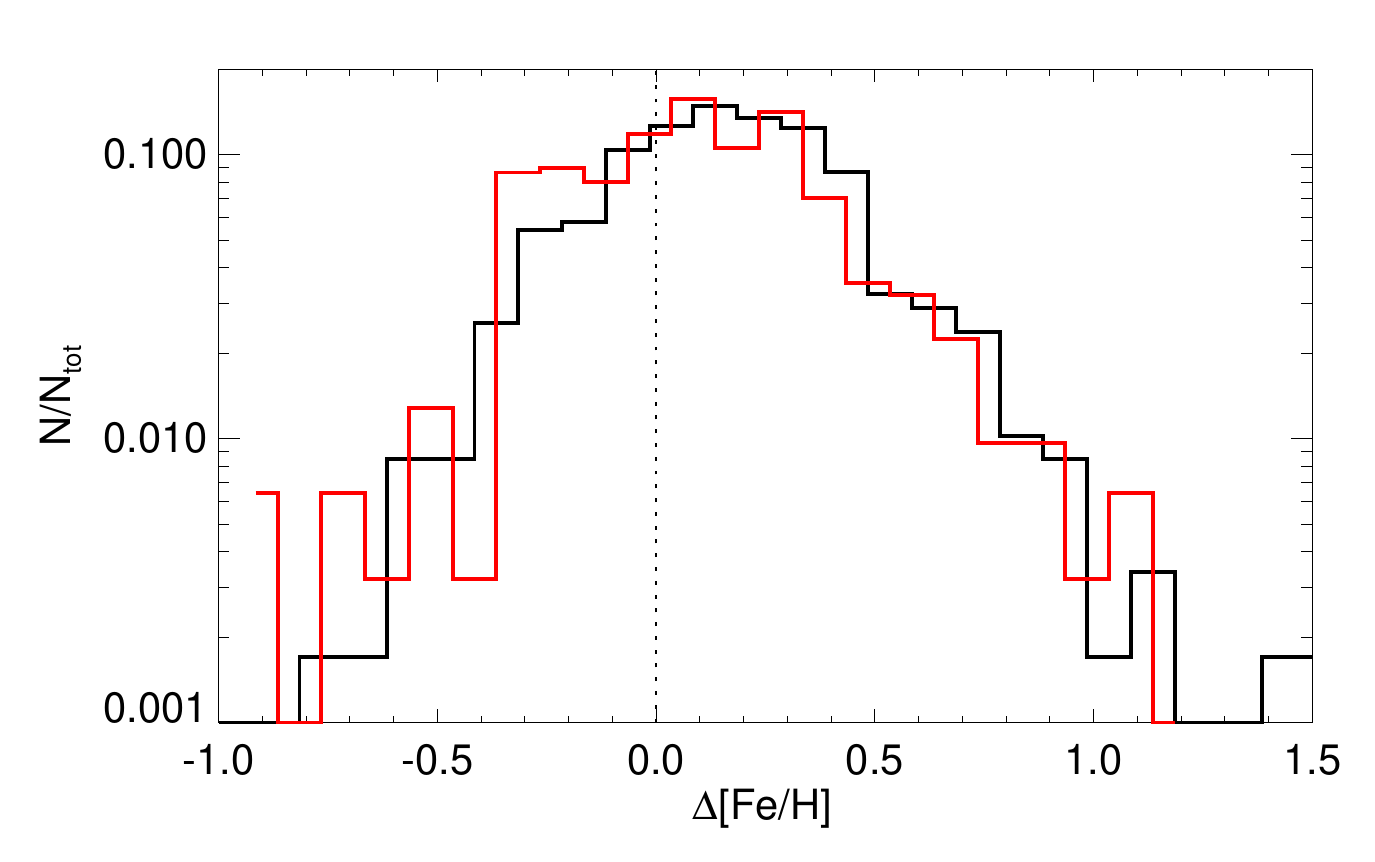}
   \includegraphics[width=\columnwidth]{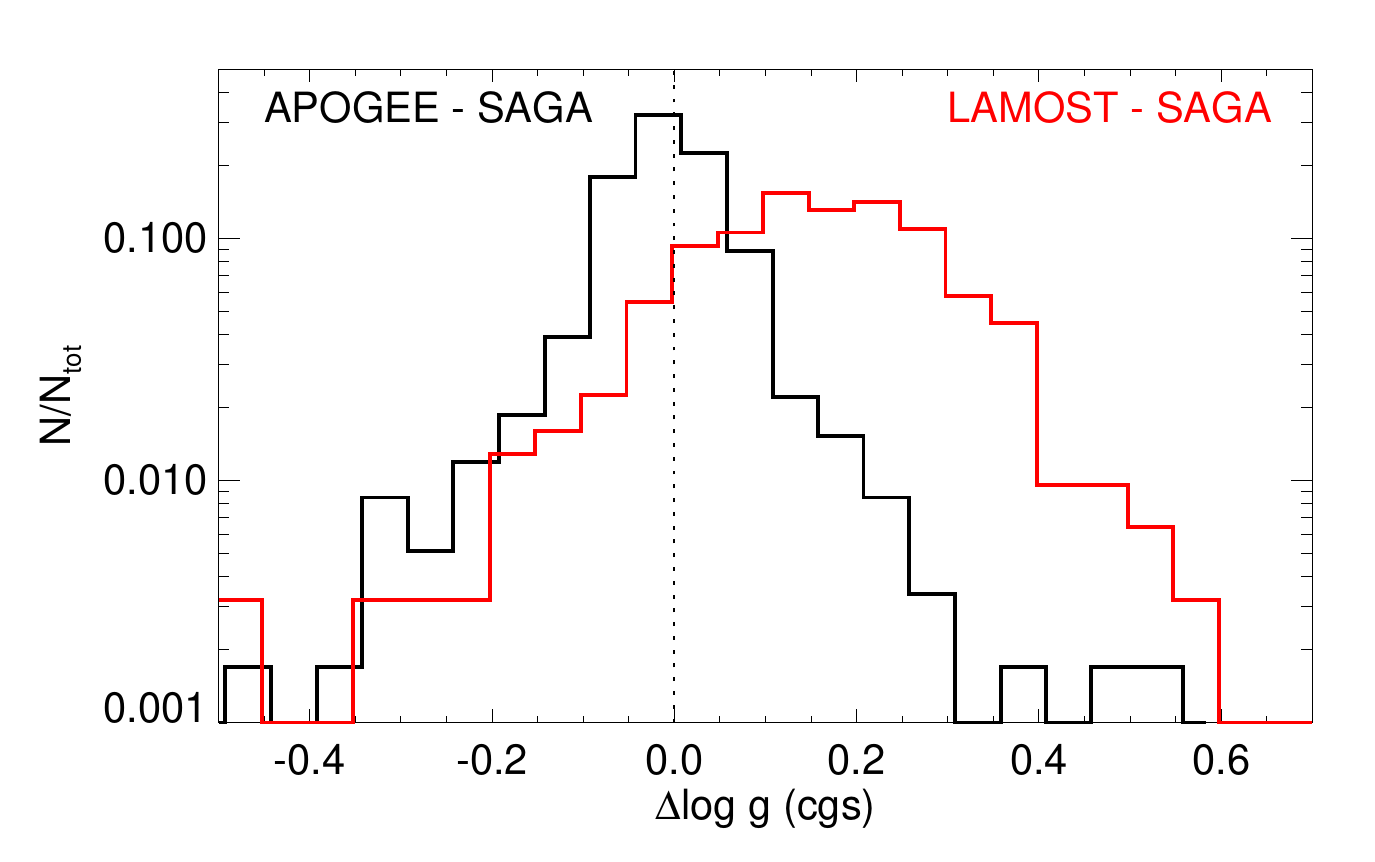} 
   \caption{Histograms of discrepancies for the main stellar atmosphere parameters between the surveys APOGEE - SAGA (black lines), and LAMOST - SAGA (red lines). See text for more details.}
  \label{fig:SAGA_APO_LAMO}
\end{figure}

\begin{figure}
  \centering  
  \includegraphics[width=1.\columnwidth]{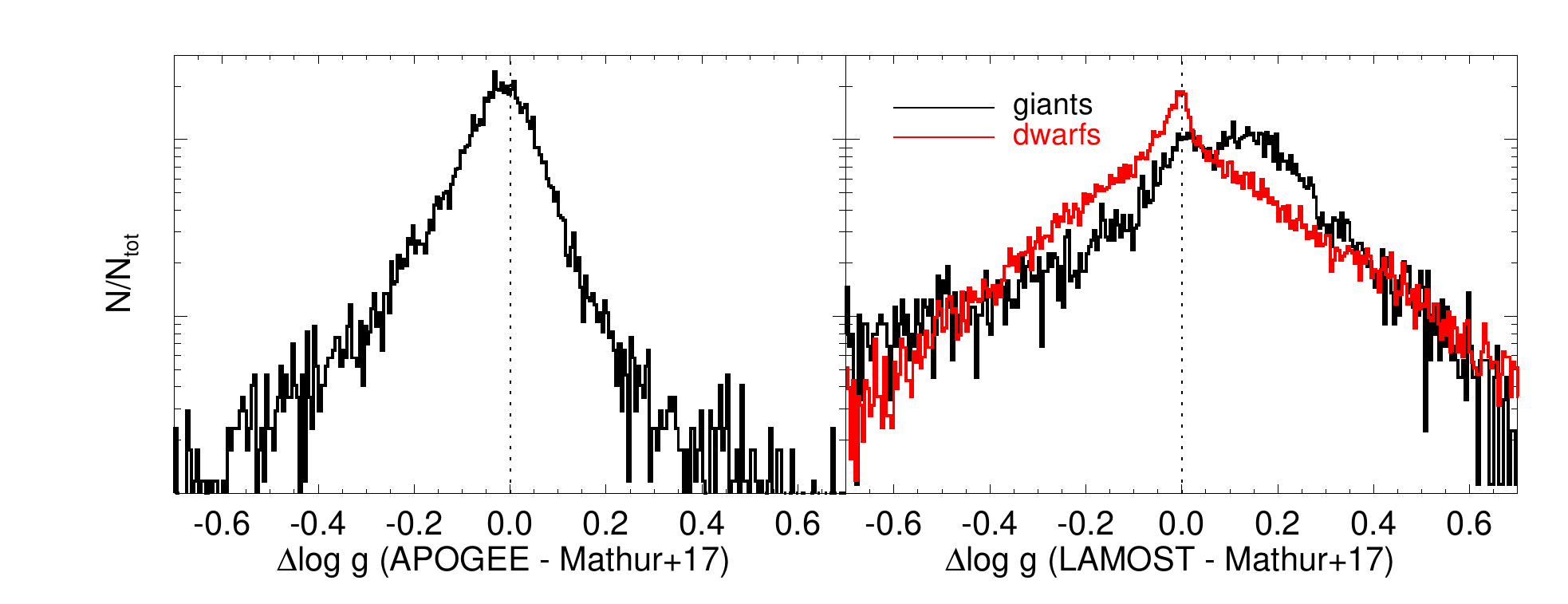}
   \caption{Discrepancies between the surface gravity derived in APOGEE, LAMOST and \cite{2017ApJS..229...30M}. See text for details.}
  \label{fig:logg_discre}
\end{figure}

\begin{table*}[htp]
\caption{Average discrepancies and standard deviations in the main stellar parameters for stars in common between APOGEE - SAGA (588 stars), and LAMOST - SAGA (312 stars) in the \emph{Kepler} field.}
\begin{center}
\begin{tabular}{ccccccc}
\hline
 & $<\Delta T_{\rm eff}>$ & $\sigma_{T_{\rm eff}}$  & $<\Delta[Fe/H]>$ & $\sigma_{\rm [Fe/H]}$ & $<\Delta log g>$ & $\sigma_{\rm log g}$  \\
\hline
\hline

APOGEE - SAGA &  --43.67 &  79.17  &  0.16 &  0.28 &  --0.01 &  0.09 \\ 
\hline 
\hline
LAMOST - SAGA &  --43.41 &    82.56  &    0.10  & 0.31  &  0.14 & 0.15  \\

\end{tabular}
\end{center}
\label{tab:APOSAGA}
\end{table*}%

In Section~\ref{sp} we reported a substantial offset in the surface gravity of 0.14 dex between the two surveys. We can make use of the NASA \emph{Kepler} mission \cite{2010Sci...327..977B} to independently check the source of the discrepancies. The \emph{Kepler} mission has obtained high-precision photometric data over the past four years. \emph{Kepler} data enable precise stellar astrophysics thanks to asteroseismic measures of red giants, where uncertainties in surface gravity $<$ 0.05 dex can be derived (\cite{2017ApJS..229...30M} and references therein). In the APOGEE-LAMOST stellar catalog there are 6846 stars in the \emph{Kepler} field.  

The Str\"omgren Survey for Asteroseismology and Galactic Archaeology (SAGA) also mapped the \emph{Kepler} field using the $uvby$ Str\"omgren photometric system. Effective temperatures in the SAGA survey are derived using the Infrared Flux Method (IRFM). This technique uses multiband optical and infrared photometry to recover the bolometric and infrared flux of each star, from which its T$_{\rm eff}$ and angular diameter can be computed \cite{2014ApJ...787..110C}. The APOGEE survey contains 588 stars in common with SAGA, while LAMOST contains 312 objects. Figure~\ref{fig:SAGA_APO_LAMO} shows the histogram of discrepancies between APOGEE and SAGA for T$_{\rm eff}$, [Fe/H] and log g in black lines, and between LAMOST and SAGA (red lines). We find a small offset of $-$43 K in T$_{\rm eff}$ between APOGEE and SAGA and also between LAMOST and SAGA. This similarity between APOGEE and LAMOST T$_{\rm eff}$ is consistent with our earlier assessment finding in Figure~\ref{fig:histo_discre}.  

There is also a consistent offset in [Fe/H] of +0.16 dex between APOGEE and SAGA and LAMOST and SAGA, together with a substantial scatter of 0.3 dex. For the surface gravities we find a good agreement with a scatter of 0.09 dex between APOGEE and SAGA. The discrepancies between LAMOST and SAGA show a clear offset in gravities of 0.14 dex, where the comparison is dominated by red giants. In Section~\ref{sp} we discussed how APOGEE surface gravities from spectroscopy are calibrated to seismic gravities, while the SAGA survey also uses global oscillation parameters to obtain log g. We summarize our findings in Table~\ref{tab:APOSAGA}. 	

\begin{table}[htp]
\caption{Surface gravity average discrepancies and its standard deviation with respect APOGEE - \cite{2017ApJS..229...30M}, and LAMOST - \cite{2017ApJS..229...30M} in the NASA \emph{Kepler} field.}
\begin{center}
\begin{tabular}{ccc}
\hline
 &$<\Delta log g>$ & $\sigma_{\rm log g}$  \\
\hline
\hline

APOGEE - Mathur17 & --0.03 &  0.13 \\ 
\hline 
\hline
LAMOST - Mathur17 (dwarfs/giants) &  0.01/0.10 & 0.23/0.24  \\

\end{tabular}
\end{center}
\label{tab:APOMA}
\end{table}%

Recently, \cite{2017ApJS..229...30M} revised stellar properties for a total number of 197,096 \emph{Kepler} targets, where the priority list for input surface gravity comes from asteroseismology. We have 17,131 stars in common between APOGEE and \cite{2017ApJS..229...30M}. With LAMOST there are 32,547 objects in common with Mathur et al. We represent in Figure~\ref{fig:logg_discre} the histogram of discrepancies for the surface gravity between APOGEE giants and LAMOST giants and dwarfs with respect to the gravities calibrated in the latter study. Additionally, in the LAMOST comparison we divided the sample in giants and dwarfs using a simple selection in log g, where stars with log g $<$ 3.5 are considered to be giants. We find good agreement between log g in APOGEE and the surface gravities revised in Mathur et al. Interestingly, the surface gravity for dwarfs stars in LAMOST show an agreement with the gravities derived in Mathur et al. for the \emph{Kepler} field, but for giants there is a clear offset and a large scatter of 0.24 dex. Calibrating LAMOST to \emph{Kepler} could resolve that discrepancy. See Table~\ref{tab:APOMA} for simple statistics describing this comparison. 

\section{\emph{The Cannon} on LAMOST}

\begin{figure}
  \centering  
  \includegraphics[width=\columnwidth]{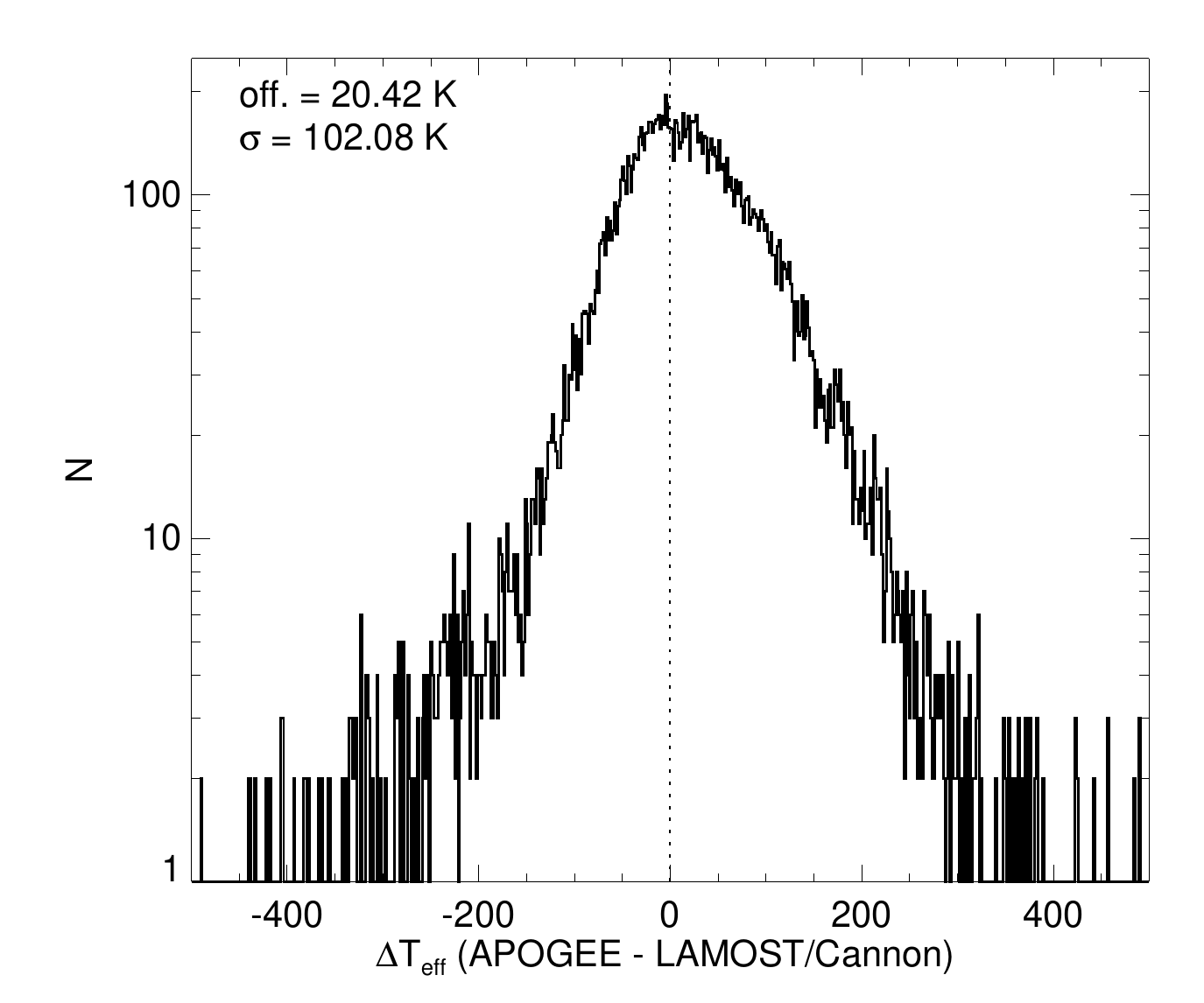}
   \includegraphics[width=\columnwidth]{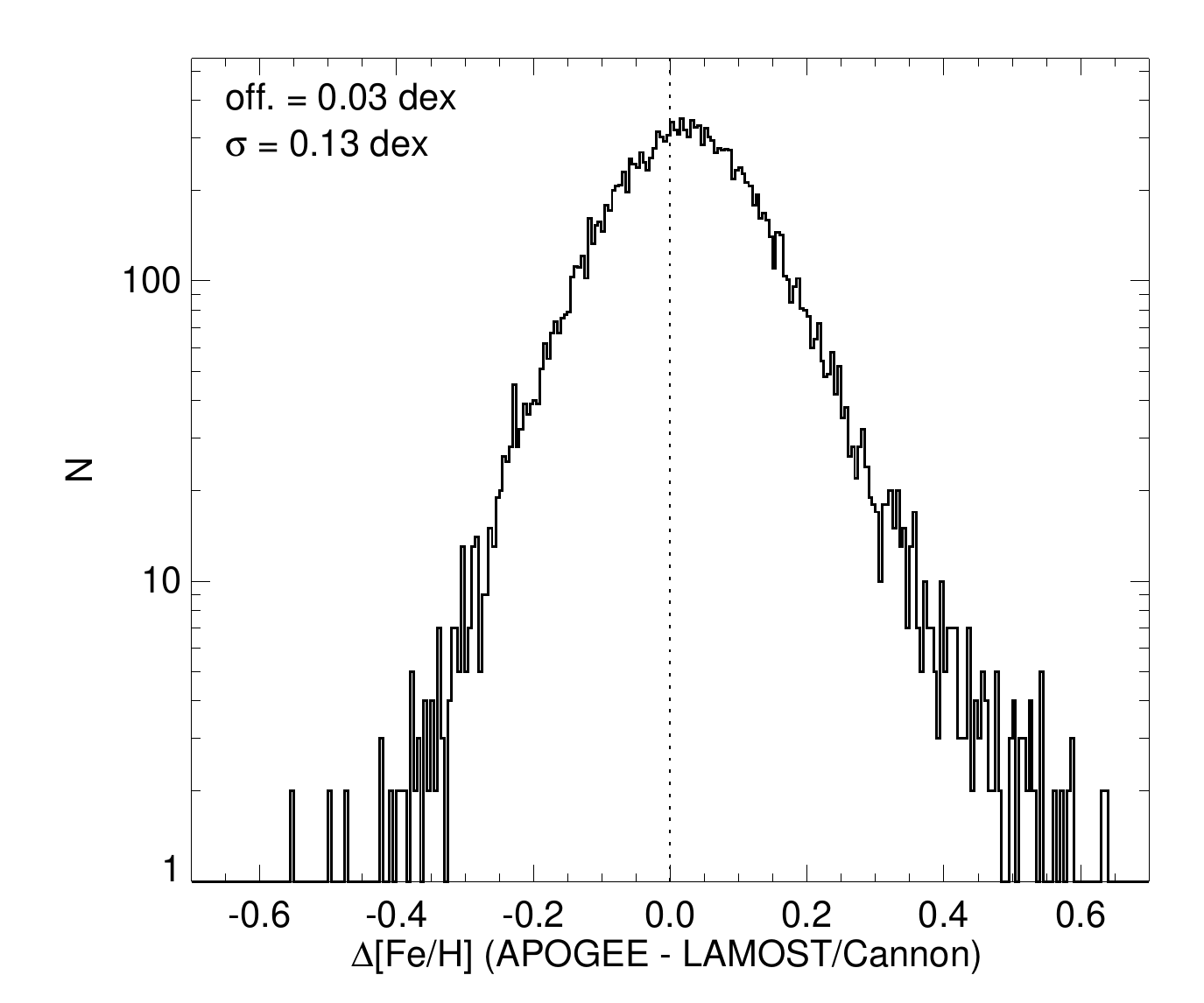}
   \includegraphics[width=\columnwidth]{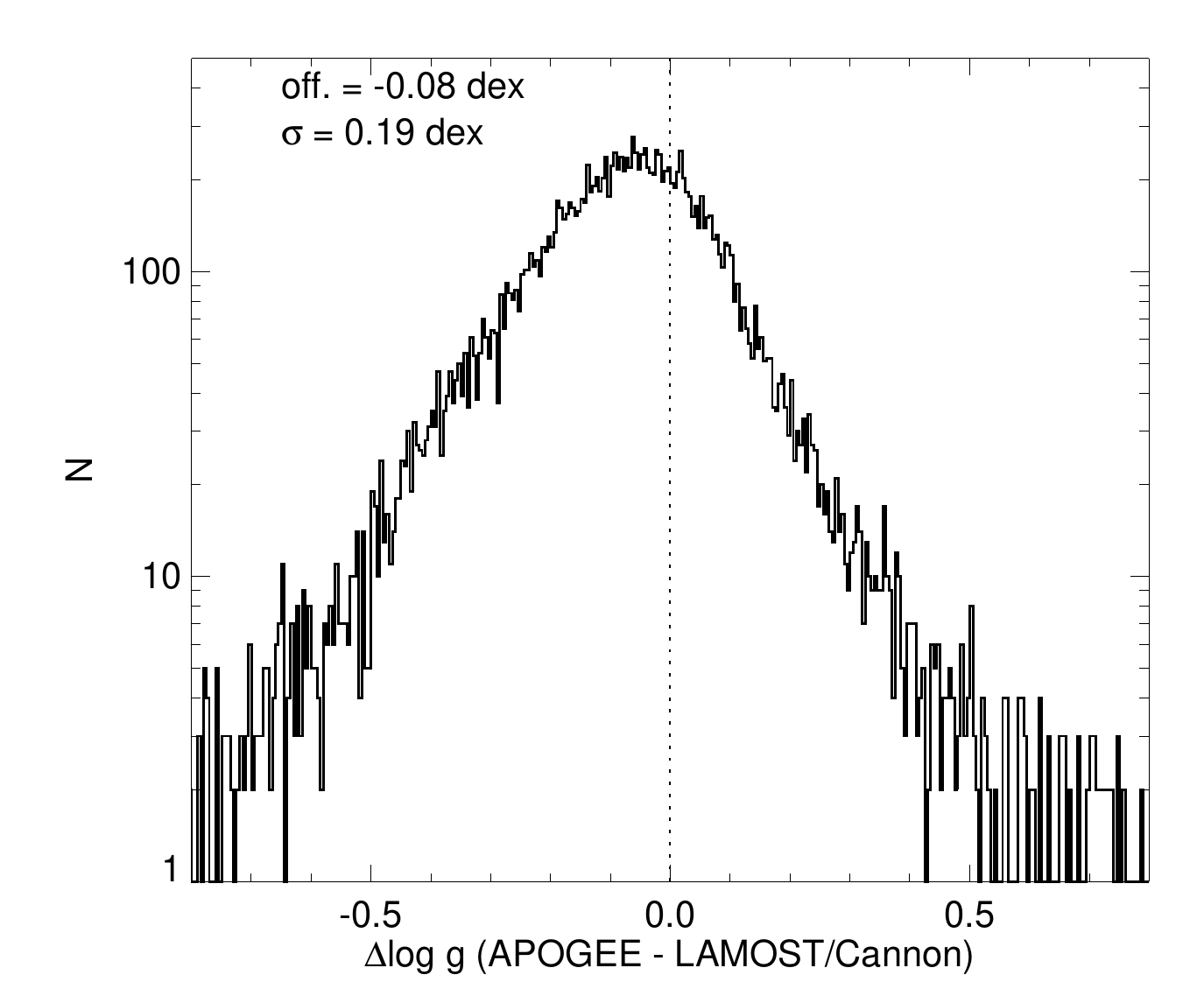} 
   \caption{Histograms of discrepancies for the main stellar atmosphere parameters between the APOGEE DR14 and LAMOST tied to APOGEE DR12 via \emph{the Cannon}. See text for more details.}
  \label{fig:APO_LAMO_CA}
\end{figure}

\cite{2017ApJ...836....5H} used a data-driven approach to spectral modeling, \emph{the Cannon} \cite{2015ApJ...808...16N}, to transfer the calibrating information from the APOGEE survey to determine precise T$_{\rm eff}$, [Fe/H] and log g, from the spectra of 454,180 LAMOST DR2 giants. \cite{2017ApJ...836....5H} select 9,952 objects common to LAMOST DR2 and APOGEE DR12 for their \emph{training set} and these objects span a representative range of parameter space for giants (see their Fig. 4 for details), but are relatively concentrated compared to the parameter ranges explored in this paper so far. We find 17,482 stars in common between APOGEE DR14 and the catalog from \cite{2017ApJ...836....5H}. Figure~\ref{fig:APO_LAMO_CA} shows the histogram of discrepancies between APOGEE DR14 and LAMOST/Cannon for T$_{\rm eff}$, [Fe/H] and log g, respectively. There is a generally good agreement between the two data-sets. The offset in [Fe/H] we found in Figure~\ref{fig:histo_discre} comparing APOGEE DR14 and LAMOST DR3 is less evident in this new comparison, however the scatter remains similar. Also, the offset in log g is still present, with a scatter of $\sim$ 0.2 dex. 

\begin{figure*}
  \centering  
  \includegraphics[width=2.\columnwidth]{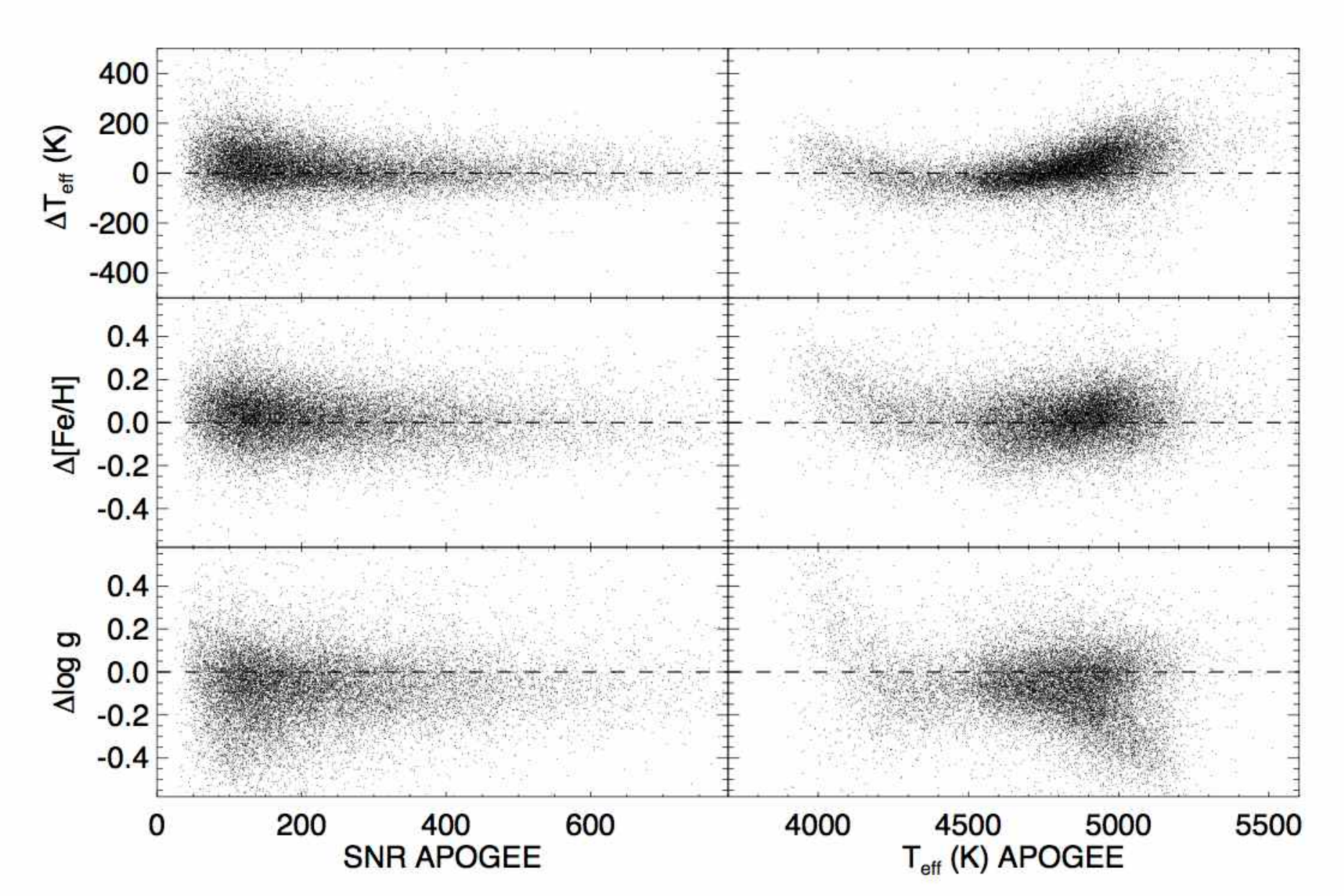}
   \includegraphics[width=2.\columnwidth]{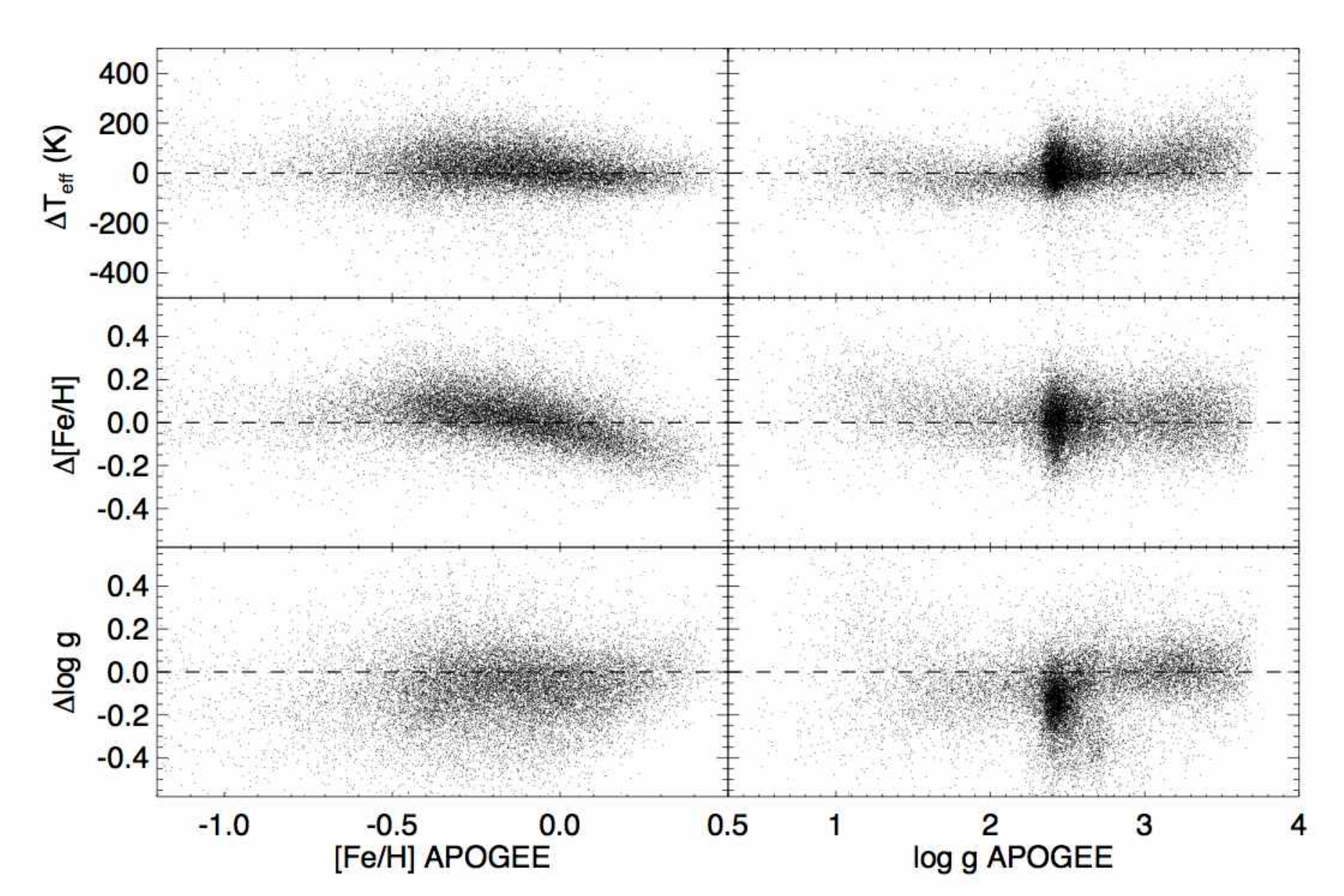}
   \caption{Main stellar parameters discrepancies between APOGEE DR14 and LAMOST, tied to APOGEE DR12 via \emph{the Cannon}, as a function of $<$SNR$>$, $<T_{\rm eff}>$, $<$[Fe/H]$>$ and $<$log g$>$ for stars in common.}
  \label{fig:APO_LAMO_CAN_trends}
\end{figure*}

We also study the behaviour of the discrepancies in the atmosphere parameters with respect to SNR, T$_{\rm eff}$, [Fe/H] and log g. From Figure~\ref{fig:APO_LAMO_CAN_trends} we learn that the discrepancies in the stellar parameters are nearly independent of the SNR of the APOGEE spectra. However, we find some dependence in the discrepancies of the stellar parameters with respect to T$_{\rm eff}$, particularly at the coolest T$_{\rm eff}$. For metal-rich stars, we also find a different trend for the [Fe/H] discrepancies with respect to [Fe/H] in APOGEE DR14. 

\begin{figure}
  \centering  
  \includegraphics[width=\columnwidth]{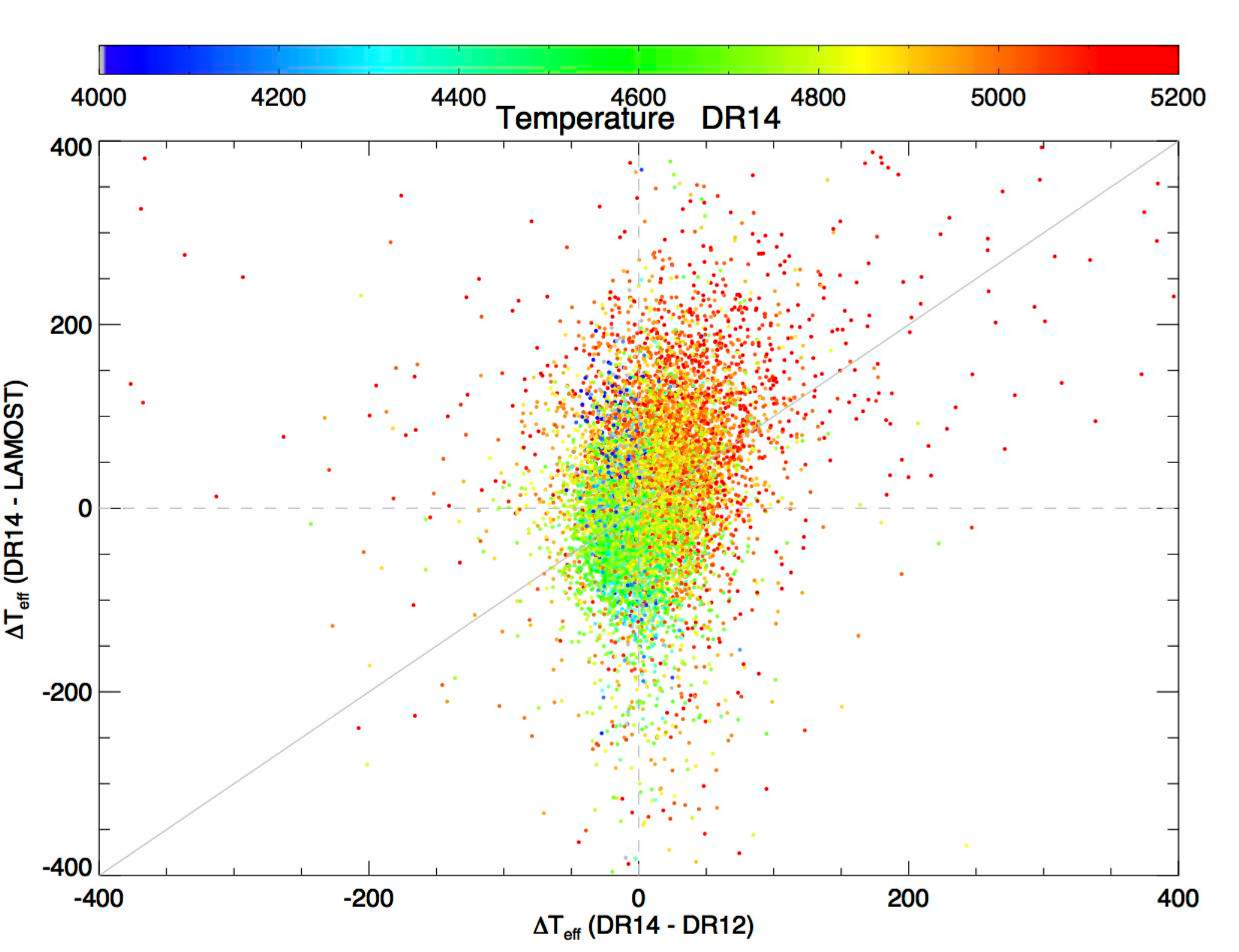}
   \includegraphics[width=\columnwidth]{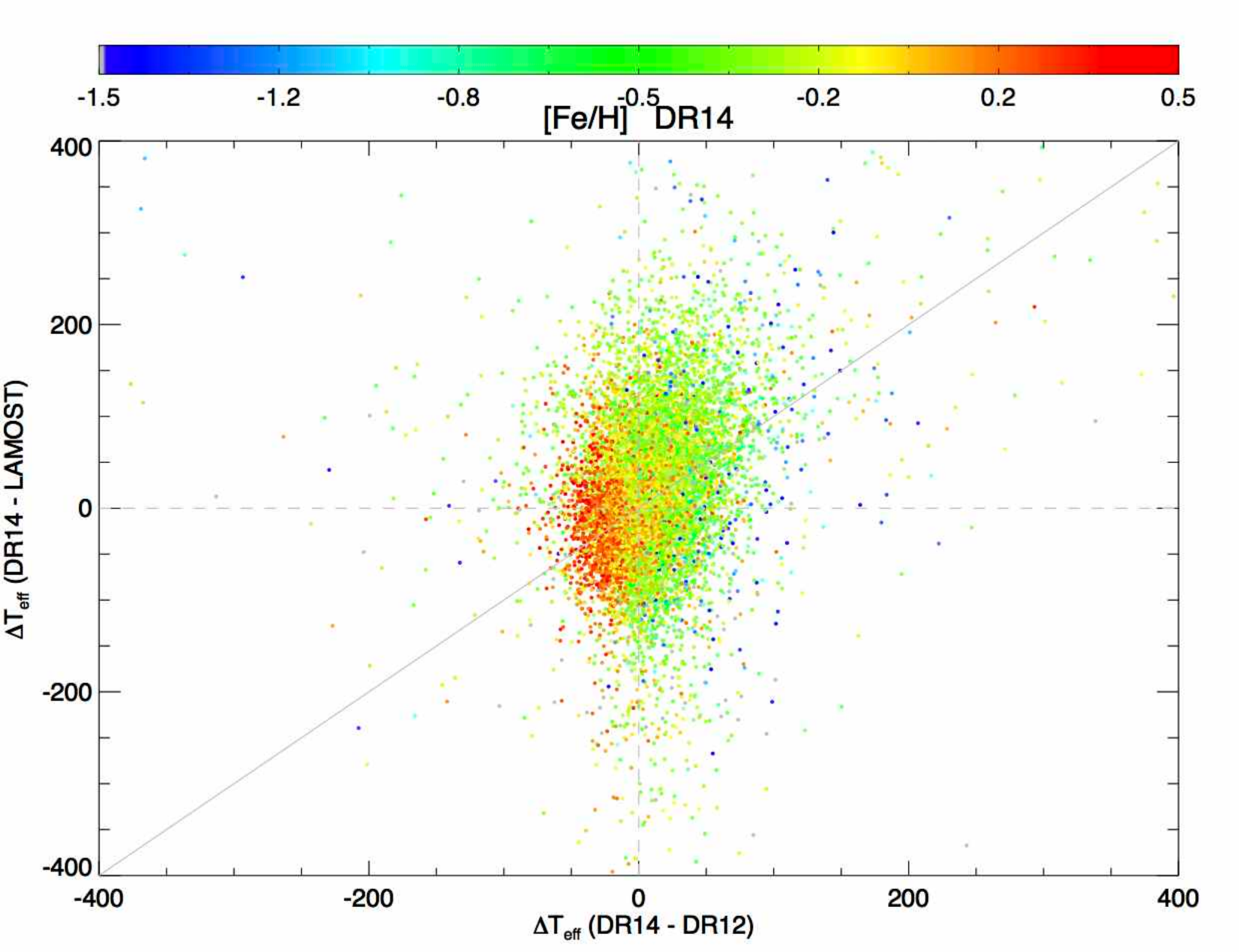}
   \includegraphics[width=\columnwidth]{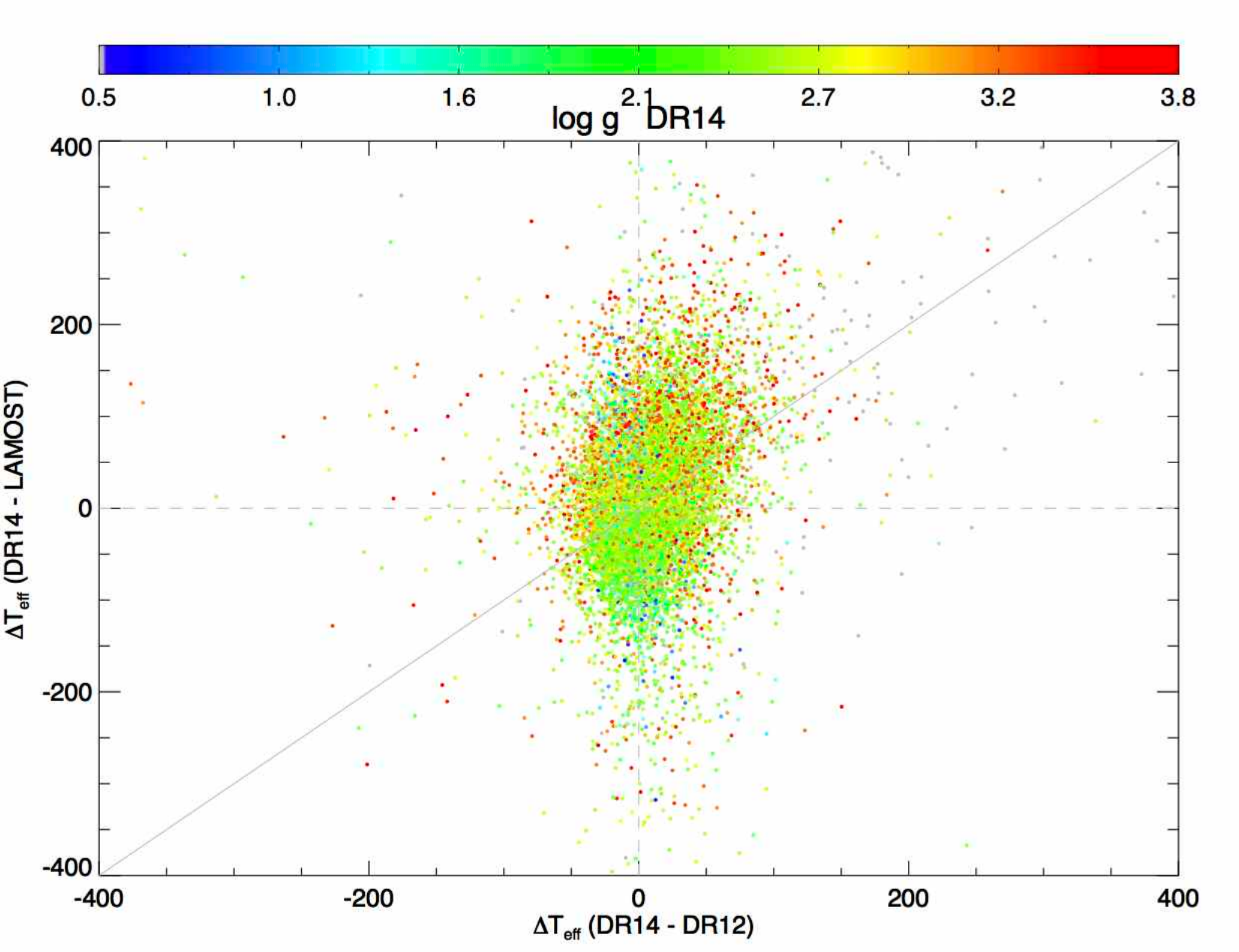} 
   \caption{T$_{\rm eff}$ discrepancies between APOGEE DR14 and DR12 with respect to APOGEE DR14 and LAMOST on the APOGEE DR12 stellar parameters space. The three graphics are color-coded by T$_{\rm eff}$, [Fe/H] and log g from APOGEE DR14.}
  \label{DR14_DR12_LA_Teff}
\end{figure}

A helpful exercise to understand how the different calibrations and data analysis pipelines are performing is to compare the stellar parameters from APOGEE DR12 to APOGEE DR14 versus LAMOST tied to APOGEE DR12 using \emph{the Cannon}. Figure~\ref{DR14_DR12_LA_Teff} shows the discrepancies in T$_{\rm eff}$ between APOGEE DR14 - APOGEE DR12 versus APOGEE DR14 - LAMOST, where the latter has been calibrated using stellar parameters from APOGEE DR12 as a reference. The figures are color-coded with respect to T$_{\rm eff}$ (top panels), [Fe/H] (middle panels) and log g (bottom panels) from APOGEE DR14, respectively. The trends in Figure~\ref{DR14_DR12_LA_Teff} show that there is a temperature and metallicity dependence in the calibration from DR12 to DR14 (Holtzman et al. in prep.). These dependences are also present in the transformation of LAMOST to APOGEE DR12 parameters space via \emph{the Cannon}, these temperature and metallicity dependences are weaker than those observed in the transformation from DR12 to DR14. Interestingly, there is no correlation between the DR14 - DR12 discrepancies and DR14 - LAMOST tied to to DR12 for T$_{\rm eff}$, where a correlation would be anticipated if the parameters derived from LAMOST spectra were brought onto the APOGEE DR12 scale. More specifically, if the LAMOST parameters and abundances were perfectly tied to the APOGEE DR12 scale, we would expect to see the known differences between DR12 and DR14 imprinted on the LAMOST parameters.

\begin{figure}
  \centering  
  \includegraphics[width=\columnwidth]{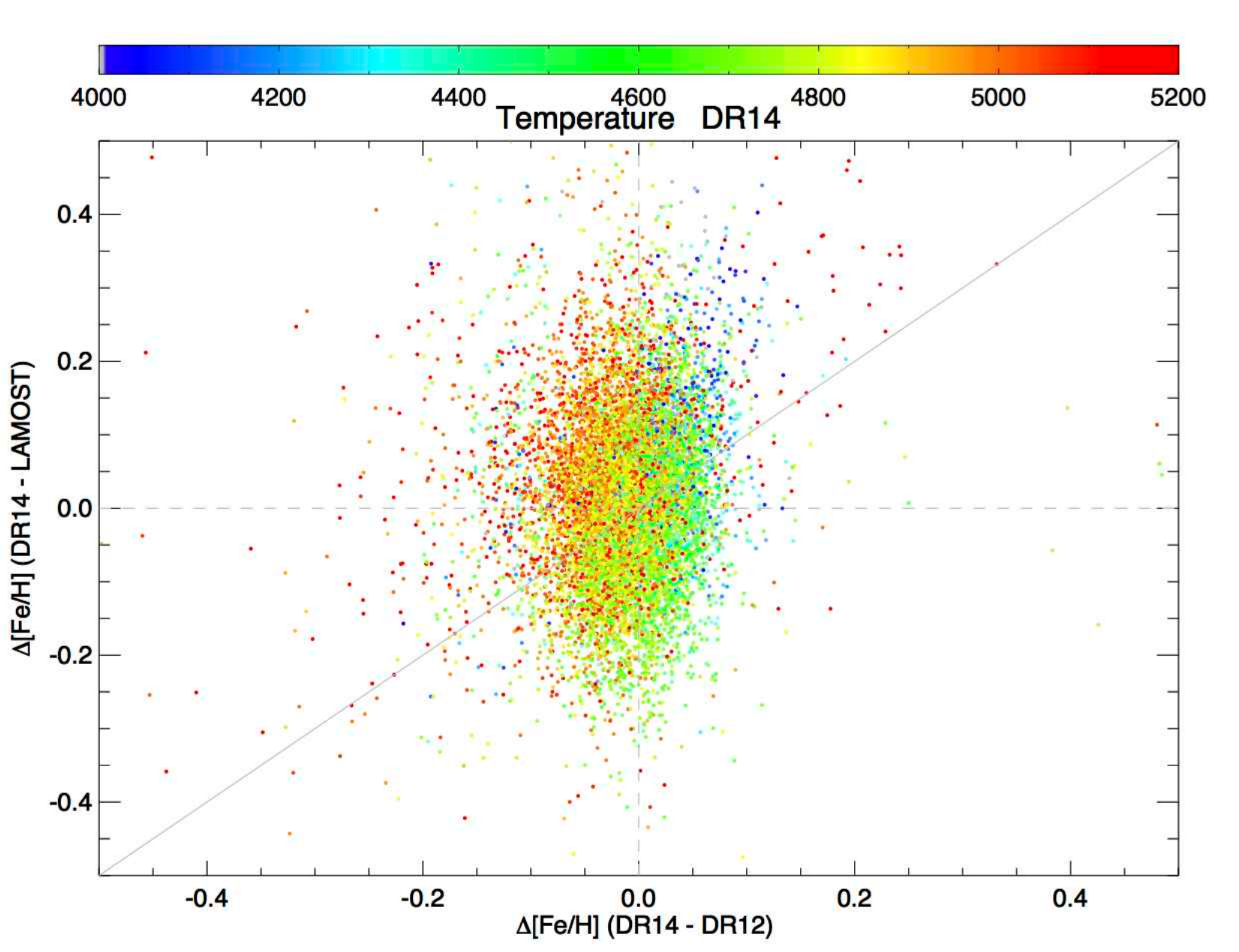}
   \includegraphics[width=\columnwidth]{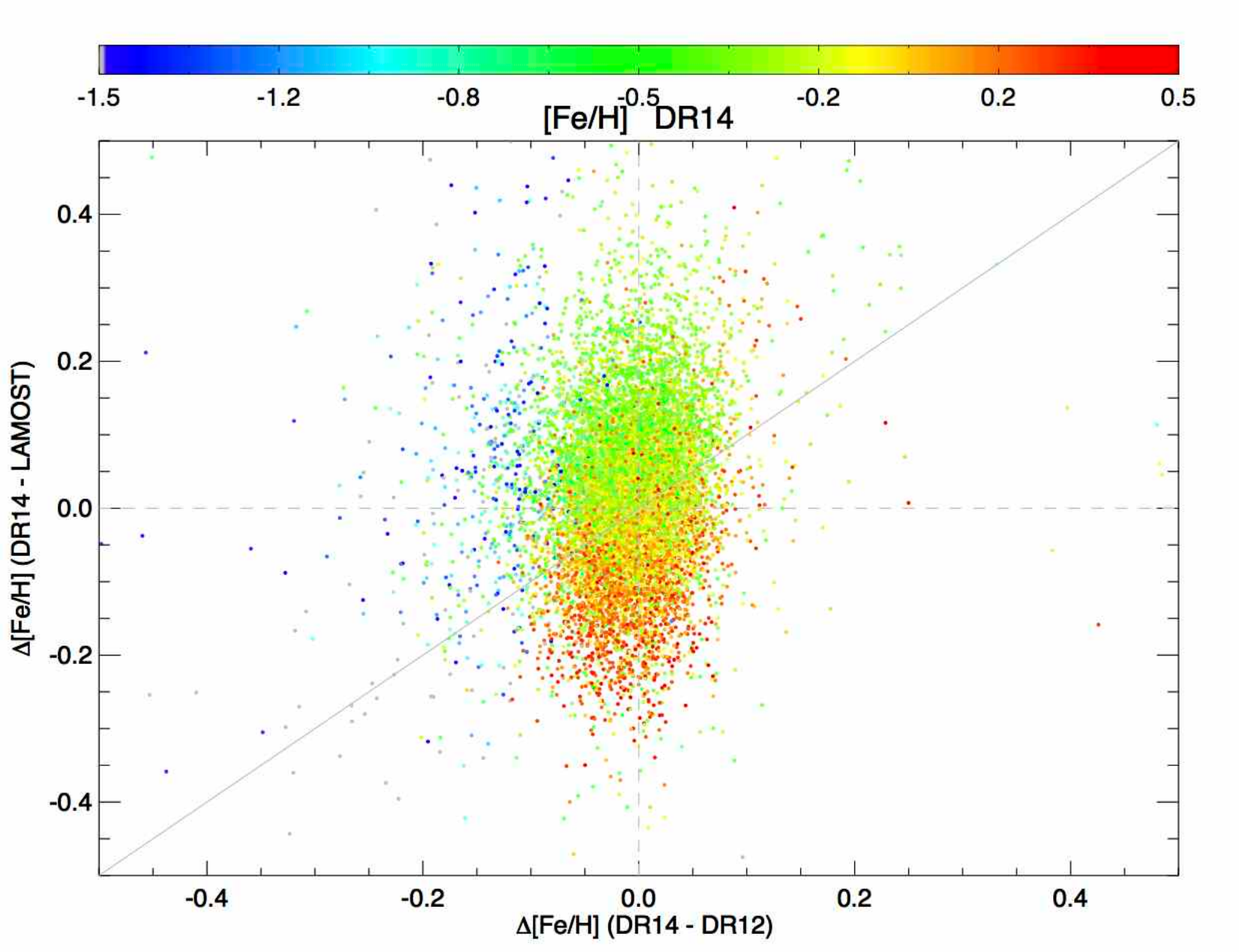}
   \includegraphics[width=\columnwidth]{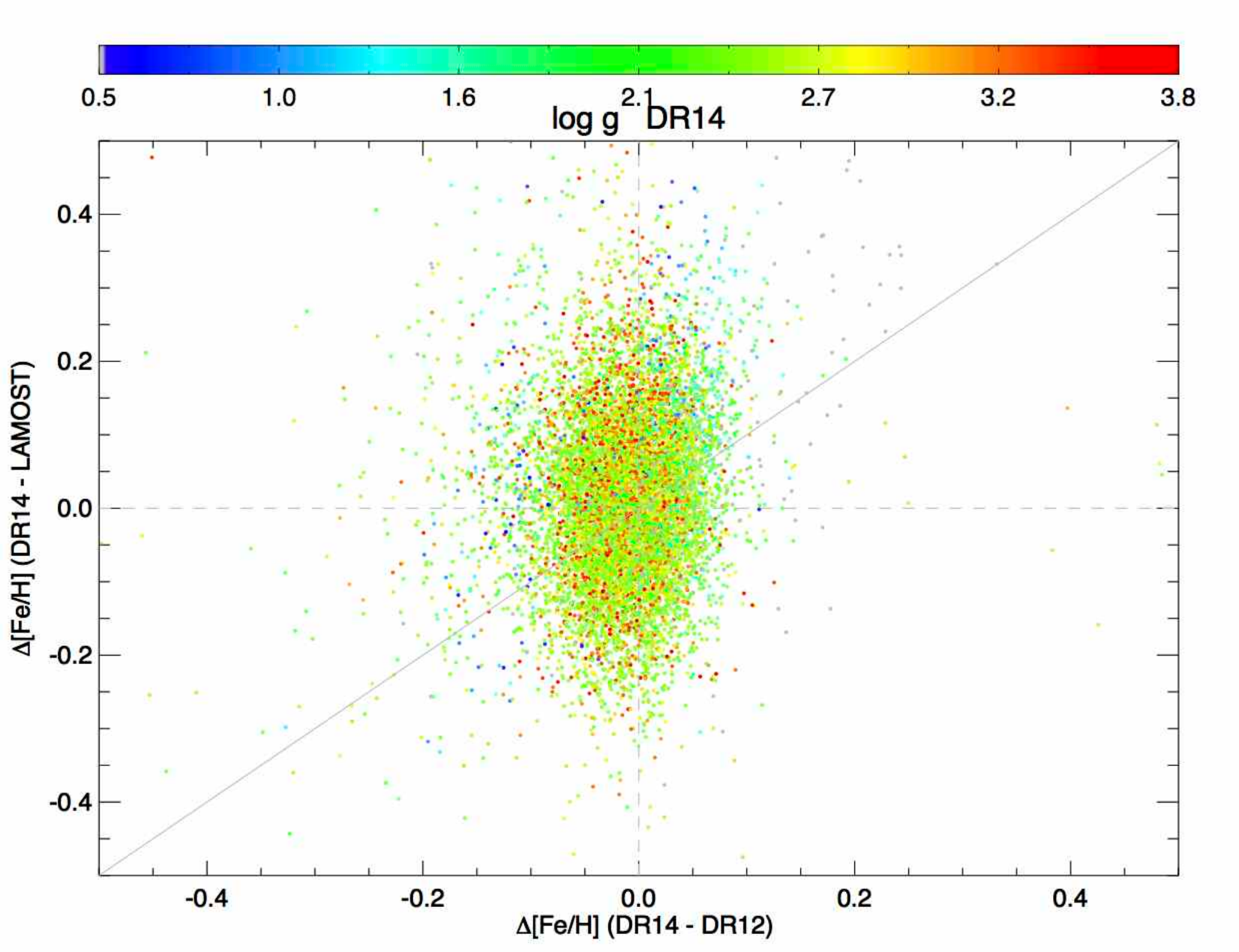} 
   \caption{[Fe/H] discrepancies between APOGEE DR14 and DR12 with respect to APOGEE DR14 and LAMOST on the APOGEE DR12 stellar parameters space. The three graphics are color-coded by T$_{\rm eff}$, [Fe/H] and log g from APOGEE DR14.}
  \label{DR14_DR12_LA_meta}
\end{figure}

In Figure~\ref{DR14_DR12_LA_meta} we have the [Fe/H] discrepancies from APOGEE DR14 - DR12 and DR14 - LAMOST calibrated by DR12 via \emph{the Cannon}. We also find no correlation between the [Fe/H] discrepancies, suggesting that LAMOST/Cannon is not entirely in the APOGEE DR12 abundances space. The graphics color-coded by T$_{\rm eff}$, [Fe/H] and log g from DR14 also indicate that there is an [Fe/H] dependence in the calibration from APOGEE DR12 to DR14.  

\begin{figure}
  \centering  
  \includegraphics[width=\columnwidth]{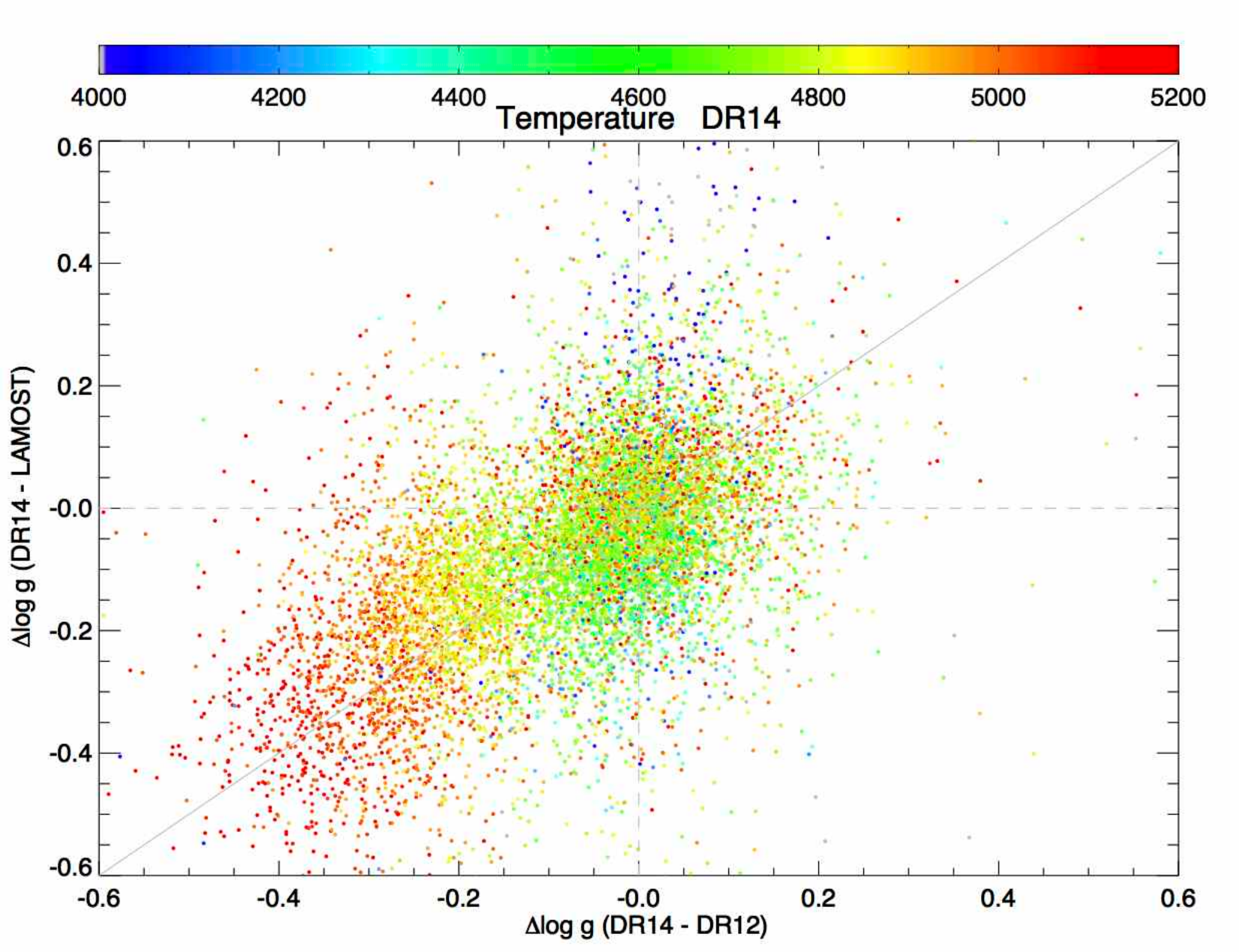}
   \includegraphics[width=\columnwidth]{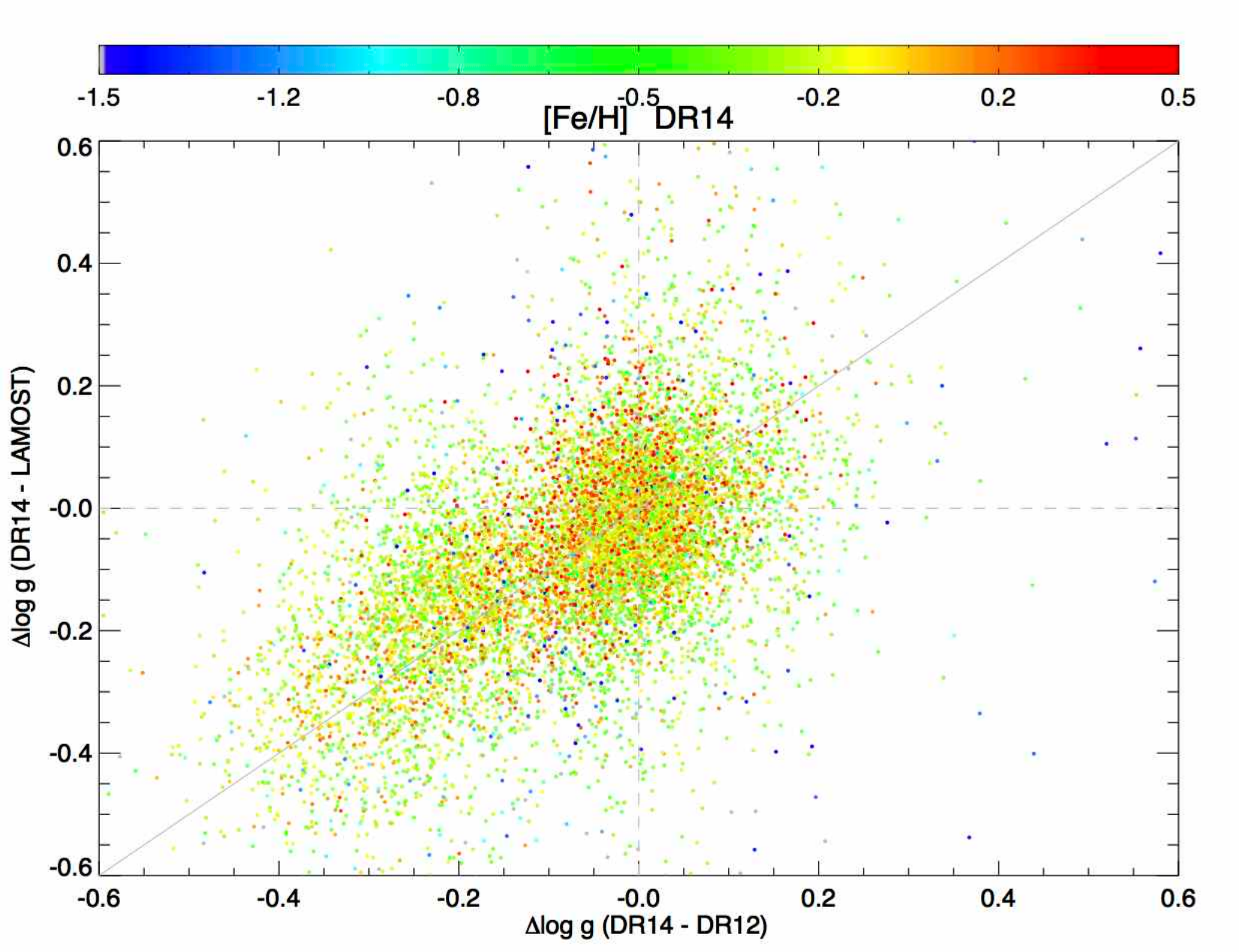}
   \includegraphics[width=\columnwidth]{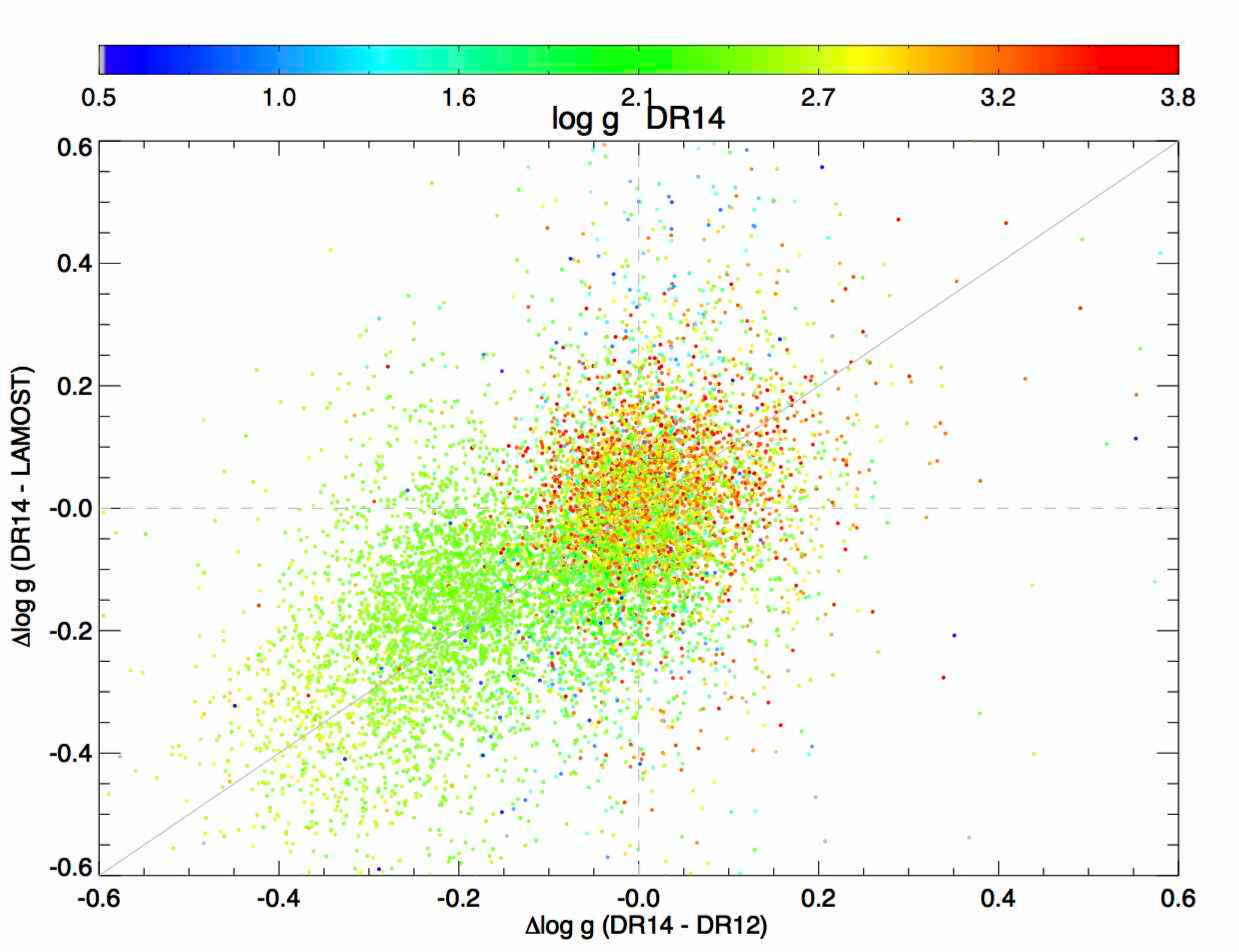} 
   \caption{Surface gravity discrepancies between APOGEE DR14 and DR12 with respect to APOGEE DR14 and LAMOST on the APOGEE DR12 stellar parameters space. The three graphics are color-coded by T$_{\rm eff}$, [Fe/H] and log g from APOGEE DR14.}
  \label{DR14_DR12_LA_grav}
\end{figure}

Finally, Figure~\ref{DR14_DR12_LA_grav} shows the surface gravity discrepancies. Unlike the previous comparison, we do see a weak correlation that would suggest the mapping of LAMOST parameters to APOGEE DR12. The correlation has a broad scatter and there is a stellar temperature dependency (top panel in Fig.~\ref{DR14_DR12_LA_grav}). We do not observe an [Fe/H] dependence in the gravity discrepancies. 

\section{Conclusions}

In this section we summarize our findings from this exercise.

   \begin{enumerate}
      \item We created the APOGEE-LAMOST stellar catalog by combining APOGEE DR14 and LAMOST DR3, and find a total of 42,420 stars in common. Most of the common stars lie in the Galactic anti-center and in the North Galactic Cap. There are also common objects in the \emph{Kepler} field. The luminosity distribution in the $H$-band covers a large magnitude range, from 7 to nearly 14 mag. Most of the stars are in the magnitude range from 9 to 12 mag. We find that the APOGEE survey has 80$\%$ of these stars with SNR $>$ 100, and 40$\%$ with SNR $>$ 200, while the LAMOST survey has 40 per cent of stars with SNR $>$ 100, and 10$\%$ with a SNR $>$ 200 in the $z$-band.\\
      
      \item The histogram of discrepancies between APOGEE and LAMOST RVs shows a clear offset of 4.54 $\pm$ 0.03 km s$^{-1}$, with a scatter of 5.8 km s$^{-1}$. We observe that most of the scatter in the discrepancies comes from LAMOST RVs uncertainties, suggesting that the average LAMOST measurement error for the RVs is $\sim$ 6 km s$^{-1}$.  The median reported LAMOST RV uncertainty in APOGEE-LAMOST stellar catalog is 6.5 km s$^{-1}$; in good agreement with our findings in this comparison study. This seems to be a universal offset because no clear systematic trends are found between the RV discrepancies and the stellar parameters, except for a weak trend of amplitude 1 km s$^{-1}$ as a function of $<$[Fe/H]$>$. \\
      
      \item We also investigate histograms of discrepancies for T$_{\rm eff}$, [Fe/H] and log g between the APOGEE and LAMOST surveys for the in common objects. We remind the reader that we use \emph{calibrated} APOGEE stellar parameters, while LAMOST parameters are not calibrated. See Section~\ref{sp} for details. We observe a small offset in the effective temperature of about 13 K, with a scatter of 155 K. A small offset in [Fe/H] of about 0.06 dex together with a scatter of 0.13 dex is also found. We notice that the largest offset between both surveys occurs in the surface gravities, where a deviation of 0.14 dex is observed with a substantial scatter of 0.25 dex. The histograms of discrepancies between the measured stellar parameters in APOGEE and LAMOST are not symmetric, suggesting that other systematic effects may be present in the data. For a detailed study on the reported uncertainties we refer the reader to Section~\ref{uncer}.\\
         
      \item We use the APOGEE-LAMOST stellar catalog to explore the existence of systematic errors in the atmosphere stellar parameters. $\Delta T_{\rm eff}$ has small variations with respect to the APOGEE SNR, where the amplitude is smaller than 80 K. We observe that $\Delta T_{\rm eff}$ changes sign for stars colder than $\sim$ 4500 K and for stars hotter than $\sim$ 5200 K, where the population is dominated by dwarfs. A trend between $\Delta T_{\rm eff}$ and $<$[Fe/H]$>$ is found, the sign of $\Delta T_{\rm eff}$ flips at solar metallicity. There are no evident trends between $<$log g$>$ and $\Delta T_{\rm eff}$. The average discrepancy in [Fe/H] varies with respect to the $<$SNR$>$ and the main stellar atmosphere parameters. For example, we find a relation between $<$[Fe/H]$>$ and $\Delta$[Fe/H]. Moreover, the metallicity offset is larger for colder stars, while for stars with T$_{\rm eff}$ $>$ 5200 K, $\Delta$[Fe/H] $\sim$ 0.0 dex. $\Delta$[Fe/H] is independent of the $<$[Fe/H]$>$ for the range --0.5 $<$ [Fe/H] $<$ 0.5 dex, however there is a relation between the two quantities for stars more metal-poor than --0.5 dex. Interestingly, for stars in the surface gravity range 1.0 $<$ log g $<$ 1.5 we do not see the offset in [Fe/H]. As mentioned previously, there is an offset of $\sim$ 0.15 dex in $\Delta$log g. This offset is independent of the SNR of the observed spectra. For stars colder than 4800 K a dependence between $\Delta$log g and T$_{\rm eff}$ is observed. The parameter $\Delta$log g as a function of $<$[Fe/H]$>$ shows variations of about 0.1 dex. We also find a dependence between $\Delta$log g and $<$log g$>$ for stars with log g $<$ 2.5 dex. The reported systematic trends in T$_{\rm eff}$ are within the uncertainties, while the reported trends for cold, metal-poor giants are significant.\\ 
         
        \item \cite{2017ApJS..229...30M} revised stellar properties for a total of 197,096 \emph{Kepler} targets, where the priority list for input surface gravity comes from asteroseismology. There are 17,131 stars in common between APOGEE and \cite{2017ApJS..229...30M}, while with LAMOST there are 32,547 objects in common. We find a good agreement between log g in APOGEE and the surface gravities revised in Mathur et al. This is expected as APOGEE surface gravities are calibrated to Kepler seismic log g. The surface gravity for dwarfs stars in LAMOST show an agreement with the gravities derived in Mathur et al. for the \emph{Kepler} field, while for giants there is a clear offset and a large scatter of 0.24 dex. \\
        
        \item \cite{2017ApJ...836....5H} used \emph{the Cannon} \cite{2015ApJ...808...16N} to transfer calibrating information from the APOGEE DR12 survey to determine precise T$_{\rm eff}$, log g and [Fe/H] calibration for the spectra of 454,180 LAMOST giants. We find 17,482 stars in common between APOGEE DR14 and the catalog from \cite{2017ApJ...836....5H}. There is a general good agreement between the two data-sets. The offset in [Fe/H] we found comparing APOGEE DR14 and LAMOST DR3 is less evident in this new comparison, however the scatter remains similar. Also, the offset in log g is still present, with a scatter of $\sim$ 0.2 dex, which is similar to that found between the APOGEE gravities and these found from the LAMOST/Cannon pipeline. We learn that the discrepancies in the stellar parameters are nearly independent of the SNR in the APOGEE spectra. However, we find dependences in the discrepancies of the stellar parameters with respect to T$_{\rm eff}$. For the metal-rich stars, we also find a different trend for the [Fe/H] discrepancies with respect to [Fe/H] in APOGEE DR14. There is also a temperature and metallicity dependence in the calibration from DR12 to DR14. These dependences are also present in the transformation of LAMOST to APOGEE parameters space via \emph{the Cannon}, however these temperatures and metallicity dependences are weaker than these observed in the transformation from DR12 to DR14. We report no correlation between the DR14 - DR12 discrepancies and DR14 - LAMOST tied to to DR12 for T$_{\rm eff}$, where a correlation should be expected if LAMOST is transferred to APOGEE DR12 correctly. We also find no correlation between the [Fe/H] discrepancies, suggesting that LAMOST/Cannon is not entirely in the APOGEE DR12 stellar parameters space, or due to LAMOST low resolution the uncertainties in the stellar parameters are large. We also report a [Fe/H] dependence in the calibration applied in the stellar parameters from APOGEE DR12 to DR14. We find a weak correlation between APOGEE DR14 - DR12 and LAMOST on DR12 surface gravity for stars hotter than 4800 K and in the log g range from 2.0 and 2.8 dex, however we do not observe an [Fe/H] dependency in the gravity discrepancies. This correlation can be explained because both catalogs use seismic gravities from \emph{Kepler} to calibrate the surface gravity. 	
                 
   \end{enumerate}

\begin{acknowledgements}

The authors thank the anonymous referee for the useful comments and suggestions. BA and SRM acknowledge support from National Science Foundation grant AST-1616636. DAGH acknowledges support provided by the Spanish Ministry of Economy and Competitiveness (MINECO) under grant AYA-2017-88254-P. HJ acknowledges support from the Crafoord Foundation and Stiftelsen Olle Engkvist Byggm\"astare. Support for this work was provided by NASA through Hubble Fellowship grant 51386.01 awarded to R.L.B. by the Space Telescope Science Institute, which is operated by the Association of  Universities for Research in Astronomy, Inc., for NASA, under contract NAS 5-26555. SzM has been supported by the Premium Postdoctoral Research Program of the Hungarian Academy of Sciences, and by the Hungarian NKFI Grants K-119517 of the Hungarian National Research, Development and  Innovation Office. Funding for the Sloan Digital Sky Survey IV has been provided by the Alfred P. Sloan Foundation, the U.S. Department of Energy of Science, and the Participating
Institutions. SDSS-IV acknowledges support and
resources from the Center for High-Performance Computing
at the University of Utah. The SDSS web site is
www.sdss.org.
SDSS-IV is managed by the Astrophysical Research
Consortium for the Participating Institutions of the
SDSS Collaboration including the Brazilian Participation
Group, the Carnegie Institution for Science,
Carnegie Mellon University, the Chilean Participation
Group, the French Participation Group, Harvard-
Smithsonian Center for Astrophysics, Instituto de Astrof'sica de Canarias, The Johns Hopkins University, Kavli Institute for the Physics and Mathematics of
the Universe (IPMU) / University of Tokyo, Lawrence Berkeley National Laboratory, Leibniz-Institut f\"ur Astrophysik Potsdam (AIP), Max-Planck-Institut f\"ur Astronomie (MPIA Heidelberg), Max-Planck-Institut f\"ur Astrophysik (MPA Garching), Max-Planck-Institut f\"ur Extraterrestrische Physik (MPE), National Astronomical Observatories of China, New Mexico State University,
New York University, University of Notre Dame, Observatorio Nacional / MCTI, The Ohio State University,
Pennsylvania State University, Shanghai Astronomical Observatory, United Kingdom Participation Group,
Universidad Nacional Autonoma de Mexico, University
of Arizona, University of Colorado Boulder, University
of Oxford, University of Portsmouth, University of Utah,
University of Virginia, University of Washington, University
of Wisconsin, Vanderbilt University, and Yale
University.
      
\end{acknowledgements}

\bibliographystyle{plainnat}

\end{document}